\documentclass[reprint,amsmath,amssymb,aps,prb,]{revtex4-2}

\usepackage{fancyhdr}
\usepackage{graphicx}
\usepackage{float}
\usepackage{subfigure}
\usepackage{dcolumn}
\usepackage{amsmath}
\usepackage{bm}
\usepackage[colorlinks,linkcolor=blue,citecolor=blue,urlcolor= blue]{hyperref}
\usepackage{times}
\usepackage{MnSymbol}
\usepackage[version=4]{mhchem} 
\usepackage{color}
\usepackage{siunitx}
\usepackage[title]{appendix}
\usepackage{verbatim}

\begin{document}

\preprint{APS/123-QED}

\title{Spin-valley-polarized Weiss oscillations in monolayer 1{\it T}$^{\prime}$-\ce{MoS2}}
\author{Y. Li$^{1}$}
\author{W. Zeng$^{2}$}
\author{R. Shen$^{1,3}$}%
 \email{ E-mail: shen@nju.edu.cn}
\affiliation{%
$^{1}$National Laboratory of Solid State Microstructures and School of Physics, Nanjing University, Nanjing 210093, China\\
$^{2}$Department of Physics, Jiangsu University, Zhenjiang 212013, China\\
$^{3}$ Collaborative Innovation Center of Advanced Microstructures, Nanjing University, Nanjing 210093, China
}

\date{\today}        

\begin{abstract}

Monolayer 1{\it T}$^{\prime}$-\ce{MoS2} exhibits spin- and valley-dependent massive tilted Dirac cones with two velocity correction terms in low-energy effective Hamiltonian. We theoretically investigate the longitudinal diffusive magnetoconductivity of monolayer 1{\it T}$^{\prime}$-\ce{MoS2} by using the linear response theory. It is shown that, when the Fermi level is close to the spin-orbit coupling gap, the Weiss oscillation splits into two branches and exhibits spin-valley polarization in the presence of both a spatial periodic electric potential modulation in the lateral direction and a nonzero perpendicular electric field. The spin-valley polarization stems from the interplay between the tilted Dirac cones, the spin-orbit coupling gap, and the external electric potential modulation, and can be treated as a signature of monolayer 1{\it T}$^{\prime}$-\ce{MoS2}. When the Fermi level is far from the spin-orbit coupling gap, the spin-polarization appears in the Weiss oscillation under a magnetic field modulation in the lateral direction. This polarization behavior arises from the interplay between the tilted Dirac cones, the spin-orbit coupling, and the external magnetic field modulation, indicating that a finite spin-orbit coupling gap is not indispensable for polarization in the Weiss oscillation.

\end{abstract}

\maketitle\thispagestyle{fancy}

\section{\label{sec:level1}Introduction}

Monolayer trasition metal dichalcogenides (TMDCs) of the {\it T}$^{\prime}$ structure phase, which have the general chemical fomula \ce{MX2} with M (W, Mo) and X (Te, Se, S), was theoretically predicted to be a quantum spin Hall insulator  \cite{1} and has recently attracted considerable attention \cite{1,2,3,4,5,6,7,8,9,10,11,12}. As a typical material of monolayer 1{\it T}$^{\prime}$-TMDCs, monolayer 1{\it T}$^{\prime}$-\ce{MoS2} \cite{1} possesses anisotropic tilted bands around Dirac points with linear energy dispersions in the momentum space \cite{13,14,15} similar to that in 8-{\it Pmmn} borophene \cite{16,17}, and undergoes a topological phase transition \cite{18,19,20} induced by a perpendicular electric field such as that in silicene \cite{21}. Due to the anisotropic tilted Dirac bands, the strong spin-orbit interaction and the controllable band gap, monolayer 1{\it T}$^{\prime}$-\ce{MoS2} has already led to various intriguing findings such as anisotropic longitudinal optical conductivities \cite{22}, anisotropic plasmon excitations and static screening effects \cite{23}, and electric field modulated valley- and spin-dependent Klein tunneling \cite{24}.

The information of the electron spectrum of two-dimensional systems can be successfully obtained by the interaction of electrons with artificially created periodic modulation (electric, magnetic or both) with periods in the submicrometer range \cite{25,26,27,28,29,30}. Under such spatial periodic modulations, it is known as Weiss oscillation \cite{25,26,31}, that the longitudinal diffusive conductivity oscillates with the inverse magnetic field periodically. Weiss oscillations have been investigated in pristine graphene \cite{32,33}, bilayer graphene \cite{34}, $\alpha$-$\mathcal{T}_{3}$ lattices \cite{35}, and phosphorene \cite{36}. A beating pattern was predicted in spin-orbit-coupled two-dimensional electron gas (2DEG) \cite{37,38} and silicene \cite{39,40}. The tilting-induced valley polarization was discovered in 8-{\it Pmmn} borophene \cite{41}. However, the signature of Weiss oscillations in monolayer 1{\it T}$^{\prime}$-\ce{MoS2} has not been studied yet, where the interplay between the tilted Dirac cones and the spin-orbit coupling may bring distinct spin and/or valley polarization. In addition, whether a finite spin-orbit coupling gap is indispensable for spin-polarization in the Weiss oscillation deserves our attention.

In this work, we theoretically investigate Weiss oscillations in monolayer 1{\it T}$^{\prime}$-\ce{MoS2} in low temperature regime by using the linear response theory. When the Fermi surface is close to the spin-orbit coupling gap, we find that the Weiss oscillation is spin-valley-polarized in the presence of both a spatial periodic electric potential modulation in the lateral direction and a nonzero perpendicular electric field. This spin-valley polarization in the Weiss oscillation arises from the interplay between the tilted Dirac cones, the spin-orbit coupling gap, and the external electric potential modulation, which can be a signature of monolayer 1{\it T}$^{\prime}$-\ce{MoS2}. In fact, silicene is a material with the spin-orbit coupling gap but is lack of the tilted Dirac cones and 8-{\it Pmmn} borophene is a material with the tilted Dirac cones but is lack of the spin-orbit coupling gap. Neither of them can exhibit the spin-valley polarization in the Weiss oscillation \cite{40,41} with a single electric potential modulation. Comparing to the spin-polarized and valley-polarized Weiss oscillation in silicene \cite{40}, the tilted Dirac cones participate in the spin-valley-polarized Weiss oscillation in monolayer 1{\it T}$^{\prime}$-\ce{MoS2} instead of an additional spatial periodic electric field modulation along the $z$ direction. On the other hand, by raising the Fermi surface high enough so that the spin-orbit coupling gap can be neglected, the Weiss oscillation is spin-polarized with a magnetic field modulation. This polarization behavior arises from the interplay between the tilted Dirac cones, the spin-orbit coupling, and the external magnetic field modulation, and is different from that in silicene \cite{40}, where the spin-orbit coupling participates in the spin-polarization in the form of a finite spin-orbit coupling gap.

The rest of the paper is organized as follows. In Sec.~\ref{sec:level2}, we introduce the low-energy effective Hamiltonian of monolayer 1{\it T}$^{\prime}$-\ce{MoS2} and obtain the Landau levels and eigenfunctions. The modulation induced Weiss oscillation in the longitudinal diffusive magnetoconductivity is discussed in Sec.~\ref{sec:level3}. Finally, we conclude in Sec.~\ref{sec:level4}.

\section{\label{sec:level2}Model Hamiltonian and Landau levels}
\label{sec:2}

\subsection{\label{sec:level2A}Model Hamiltonian}

We start with the low-energy $\bm{k} \cdot \bm{p}$ Hamiltonian of the monolayer 1{\it T}$^\prime$-\ce{MoS2} in the $x$-$y$ plane with a perpendicular electric field $E_{z}$ in the $z$ direction. The Hamiltonian in the $\bm{\sigma} \otimes \bm{s}$ space reads \cite{1,24}
\begin{equation}
\label{eq. 1}
\mathcal{H} = \mathcal{H}_{\bm{k} \cdot \bm{p}} + \mathcal{H}_{E_{z}} + V,
\end{equation}where
\begin{equation}
\label{eq. 2}
\mathcal{H}_{\bm{k} \cdot \bm{p}} = \begin{pmatrix}
E_{p} & 0 & -iv_{1}\hbar q_{x} & v_{2}\hbar q_{y} \\
0 & E_{p} & v_{2}\hbar q_{y} & -iv_{1}\hbar q_{x} \\
iv_{1}\hbar q_{x} & v_{2}\hbar q_{y} & E_{d} & 0 \\
v_{2}\hbar q_{y} & iv_{1}\hbar q_{x} & 0 & E_{d}
\end{pmatrix} 
\end{equation}
and
\begin{equation}
\label{eq. 3}
\mathcal{H}_{E_{z}} = \alpha\Delta_{so}\begin{pmatrix}
0 & 0 & 1 & 0 \\
0 & 0 & 0 & 1 \\
1 & 0 & 0 & 0 \\
0 & 1 & 0 & 0
\end{pmatrix}.
\end{equation}
Here, $\bm{\sigma}$ stands for the Pauli matrix acting upon pseudospin space and $\bm{s}$ denotes the Pauli matrix acting upon real-spin space. The on-site energies of $p$ and $d$ orbitals are $E_{p}=\delta_{p} + \frac{\hbar^{2}q_{x}^{2}}{2m_{x}^{p}} + \frac{\hbar^{2}q_{y}^{2}}{2m_{y}^{p}}$ and $E_{d}=\delta_{d} + \frac{\hbar^{2}q_{x}^{2}}{2m_{x}^{d}} + \frac{\hbar^{2}q_{y}^{2}}{2m_{y}^{d}}$, respectively, where $q_{x,y}$ is the electron momentum, $\delta_{p} = 0.46 \mathrm{eV}$, $\delta_{d} = -0.20 \mathrm{eV}$, $m_{x}^{p} = -0.50 m_{0}$, $m_{y}^{p} = -0.16 m_{0}$, $m_{x}^{d} = 2.48 m_{0}$, $m_{y}^{d} = 0.37 m_{0}$, and $m_{0}$ is the free electron mass. $\delta_{p} > \delta_{d}$ at the $\Gamma$ point corresponds to the $p$-$d$ band inversion. The Fermi velocities along the $x$ and the $y$ directions are denoted as $v_{1} = 3.87 \times 10^{5} \mathrm{m/s}$ and $v_{2} = 0.46 \times 10^{5} \mathrm{m/s}$, respectively. Half of the fundamental spin-orbit coupling gap at the Dirac points is represented by $\Delta_{so} = 41.9 \mathrm{meV}$, $\alpha$ is equal to $\left| E_{z}/E_{c} \right|$ with the perpendicular electric field $E_{z}$ and the critical electric field $E_{c} = 1.42 \mathrm{V/nm^{-1}}$ for topological phase transition \cite{1}. Potential $V$ can be adjusted by gate voltage or doping.

After a proper unitary transformation
\begin{equation}
\label{eq. 4}
M = \frac{1}{\sqrt{2}}\begin{pmatrix}
1 & 0 & 1 & 0 \\
-1 & 0 & 1 & 0 \\
0 & 1 & 0 & 1 \\
0 & -1 & 0 & 1
\end{pmatrix},
\end{equation} \\
the Hamiltonian $M^{-1}\mathcal{H}M$ is written in a block-diagonal form. At two Dirac points $\Lambda = \pm(0,q_{0})$ with $q_{0} = 1.38 \mathrm{nm^{-1}}$, the diagonal term reads
\begin{equation}
\label{eq. 5}
\begin{split}
H_{s,\xi}(\bm{k}) &= \hbar k_{x}v_{1}\sigma_{y} - \hbar k_{y}(s v_{2}\sigma_{x} + \xi v_{-}\sigma_{0} + \xi v_{+}\sigma_{z}) \\
&+ \Delta_{so}(\alpha - s\xi)\sigma_{x} + V\sigma_{0},
\end{split}
\end{equation}
where $s = \pm$ labels the spin index ($\uparrow$ or $\downarrow$), $\xi = \pm$ stands for the valley index ($K$ or $K^\prime$), and $k_{x,y}$ is the momentum measured from the Dirac points. In Eq. (\ref{eq. 5}), the first two terms correspond to the kinetic energies with two velocity correction terms $v_{-} = \frac{\hbar q_{0}}{2}(-\frac{1}{m_{y}^{p}} - \frac{1}{m_{y}^{d}}) = 2.84 \times 10^{5} \mathrm{m/s}$ and $v_{+} = \frac{\hbar q_{0}}{2}(-\frac{1}{m_{y}^{p}} + \frac{1}{m_{y}^{d}}) = 7.18 \times 10^{5} \mathrm{m/s}$. The velocity correction term $v_{-}$, which is also known as the tilting velocity, describes the tilted nature of Dirac cones. The velocity correction term $v_{+}$ modifies the Fermi velocity in the $y$ direction and ensures that the Dirac cone will not be tipped over.

\subsection{\label{sec:level2B}Landau levels}

A perpendicular magnetic field $\bm{B} = B\bm{\hat{e}_{z}}$ can be introduced in the low-energy effective Hamiltonian in Eq. (\ref{eq. 5}) with the Landau gauge $\bm{A} = (0,Bx,0)$. After the Landau-Peierls substitution $\hbar\bm{k}\rightarrow\hbar\bm{k} + e\bm{A}$, one obtains 
\begin{equation}
\label{eq. 6}
\begin{split}
H_{s,\xi}(\bm{k}) &= \hbar k_{x}v_{1}\sigma_{y} - (\hbar k_{y} + eBx)(s v_{2}\sigma_{x} + \xi v_{-}\sigma_{0} + \xi v_{+}\sigma_{z}) \\
&+ \Delta_{so}(\alpha - s\xi)\sigma_{x} + V\sigma_{0}.
\end{split}
\end{equation}
Noting that the commutator $[H,k_{y}] = 0$, we can write the wave function in the ansatz $\Psi(\bm{r}) = \frac{1}{\sqrt{L_{y}}}e^{ik_{y}y}\psi(x)$ and the Hamiltonian is simplified as
\begin{equation}
\label{eq. 7}
\begin{split}
H &= \\
&\begin{pmatrix}
-\xi\hbar(v_{+} + v_{-})[\sqrt{\frac{v_{1}}{v_{2}}}X + \frac{\delta}{\sqrt{2}}] & -\hbar v_{c}[\partial_{X} + sX] \\
\hbar v_{c}[\partial_{X} - sX] & \xi\hbar(v_{+} - v_{-})[\sqrt{\frac{v_{1}}{v_{2}}}X + \frac{\delta}{\sqrt{2}}]
\end{pmatrix},
\end{split}
\end{equation}where $l = \sqrt{\hbar /eB}$ is the magnetic length, $v_{c} = \sqrt{v_{1}v_{2}}$ is the reduced velocity. The dimensionless position operator is defined as $X = \sqrt{\frac{v_{2}}{v_{1}}}\frac{x + x_{0}}{l}$ with $x_{0} = k_{y}l^{2} - \frac{\delta l}{\sqrt{2}}$ where $\frac{\delta l}{\sqrt{2}} = (s\alpha - \xi)\frac{\Delta_{so}l^{2}}{\hbar v_{2}}$ is an additional shift in the cyclotron center induced by the Rashba spin-orbit coupling.

Similar to N. M. R. Peres and E. V. Castro's algebraic solution for the graphene in the transverse electric field and the perpendicular magnetic field \cite{42}, we solve the eigenproblem $H\psi(x) = \epsilon\psi(x)$ in appendix \ref{app:ALGEBRAIC SOLUTION OF THE LANDAU LEVELS} and finally give the eigenvalues and the eigenfunctions as:
\begin{equation}
\label{eq. 8}
\begin{split}
\epsilon_{n} &= \frac{\sqrt{2}\hbar v_{c}}{l}\frac{1}{v_{2}^{2} + v_{+}^{2}}[-\frac{1}{2}\xi\delta v_{2}v_{-} \\
&+ \eta\sqrt{n\frac{(v_{2}^{2} + v_{+}^{2})(v_{2}^{2} + v_{+}^{2} -v_{-}^{2})^{3/2}}{v_{2}} + \frac{\delta^{2}}{4}v_{+}^{2}(v_{2}^{2} + v_{+}^{2} - v_{-}^{2})}], \\
&n = 0, 1, 2, \ldots,
\end{split}
\end{equation}

\begin{equation}
\label{eq. 9}
\begin{split}
\Psi_{\uparrow,\xi,\eta,n}(\bm{r}) &= \frac{1}{2L_{y}l}e^{ik_{y}y} \\
&\begin{pmatrix}
-\xi(\frac{C_{+}}{C_{+} + C_{+}^{-1}})^{1/2}\phi_{n-1}(\tilde{X}) + \eta(\frac{C_{-}}{C_{-} + C_{-}^{-1}})^{1/2}\phi_{n}(\tilde{X}) \\
(\frac{C_{+}^{-1}}{C_{+} + C_{+}^{-1}})^{1/2}\phi_{n-1}(\tilde{X}) + \eta\xi(\frac{C_{-}^{-1}}{C_{-} + C_{-}^{-1}})^{1/2}\phi_{n}(\tilde{X})
\end{pmatrix},
\end{split}
\end{equation}

\begin{equation}
\label{eq. 10}
\begin{split}
\Psi_{\downarrow,\xi,\eta,n}(\bm{r}) &= \frac{1}{2L_{y}l}e^{ik_{y}y} \\
&\begin{pmatrix}
-\xi(\frac{C_{+}^{-1}}{C_{+} + C_{+}^{-1}})^{1/2}\phi_{n-1}(\tilde{X}) + \eta(\frac{C_{-}^{-1}}{C_{-} + C_{-}^{-1}})^{1/2}\phi_{n}(\tilde{X}) \\
(\frac{C_{+}}{C_{+} + C_{+}^{-1}})^{1/2}\phi_{n-1}(\tilde{X}) + \eta\xi(\frac{C_{-}}{C_{-} + C_{-}^{-1}})^{1/2}\phi_{n}(\tilde{X})
\end{pmatrix},
\end{split}
\end{equation}with

\begin{equation}
\label{eq. 11}
\tilde{X} = \frac{1}{l}(\frac{\sqrt{v_{2}^{2} + v_{+}^{2} - v_{-}^{2}}}{v_{1}})^{1/2}(x + \tilde{x}_{0}),
\end{equation}

\begin{equation}
\label{eq. 12}
\tilde{x}_{0} =k_{y}l^{2} - \frac{\delta l}{\sqrt{2}} - \sqrt{2}\frac{\xi v_{-}\sqrt{\frac{v_{2}}{v_{1}}}\frac{\epsilon_{n}l^{2}}{\sqrt{2}\hbar} - \frac{\delta l}{2}(v_{+}^{2} - v_{-}^{2})}{v_{2}^{2} + v_{+}^{2} -v_{-}^{2}},
\end{equation}where $\eta = +1 (-1)$ is the band index, $n$ is the Landau-level index, $\phi_{n}(\tilde{X})$ is the harmonic oscillator wave function with $\phi_{-1}(\tilde{X}) = 0$, and $C_{\pm} = \frac{v_{+} - v_{-}}{\sqrt{v_{2}^{2} + v_{+}^{2} - v_{-}^{2}} \mp v_{2}}$.

\section{\label{sec:level3}Magnetoconductivity}
\label{sec:3}

In order to discuss the Weiss oscillation in monolayer 1{\it T}$^\prime$-\ce{MoS2} in low temperature regime, we assume a longitudinal current along the $y$ direction, the uniform electric field $E_{z}$, and the uniform magnetic field $B$ along the $z$ direction. We further introduce a weak static spatially periodic modulation (electric potential or magnetic field) in the $x$ direction, which can be considered as a small perturbation. The diffusive conductivity in the $y$ direction is calculated by the Kubo formula \cite{43}. Provided that the scattering processes are elastic or quasielastic, the diffusive conductivity is given by \cite{32,44,45}
\begin{equation}
\label{eq. 13}
\sigma_{yy} = \frac{\beta e^{2}}{L_{x}L_{y}}\sum_{\zeta}f(\epsilon_{\zeta})[1 - f(\epsilon_{\zeta})]\tau_{\epsilon_{\zeta}}(v_{y}^{\zeta})^{2},
\end{equation}
where $L_{x}$, $L_{y}$ are the dimensions of the sample, $\zeta$ labels the quantum number of the electron eigenstate, $f(\epsilon_{\zeta}) = \{1 + \mathrm{exp}[\beta(\epsilon_{\zeta} - \epsilon_{F})]\}^{-1}$ is the Fermi-Dirac distribution function with the Fermi energy $\epsilon_{F}$ and the inverse temperature $\beta = (k_{B}T)^{-1}$, $\tau_{\epsilon_{\zeta}}$ denotes the energy-dependent collision time and $v_{y}^{\zeta}$ represents the group velocity.

According to the position of the Fermi level, we consider two scenarios of Weiss oscillations in monolayer 1{\it T}$^\prime$-\ce{MoS2}. One is the low Fermi level case, where the Fermi level is close to the spin-orbit coupling gap and the  interplay between the tilted Dirac cones and the spin-orbit coupling can be expected. The other one is the high Fermi level case, where the Fermi level is far from the spin-orbit coupling gap and the spin-polarized Weiss oscillation without the spin-orbit coupling gap is anticipated.

 \subsection{Low Fermi level case}
\label{subsec:Low Fermi level case}

The one-dimensional static electric modulation considered is described by the small perturbation Hamiltonian $H_{e}^{\prime} = V_{e}^{\prime}\mathrm{cos}(Kx)$ with $K = \frac{2\pi}{a_{0}}$, where $V_{e}^{\prime}$ is the modulation strength and $a_{0}$ is the period of spatial modulation. Using perturbation theory, we find the first-order energy correction as
\begin{equation}
\label{eq. 14}
\begin{split}
\Delta\epsilon_{e} &= \int_{-\infty}^{\infty} \,dx\int_{0}^{L_{y}} \,dy\Psi^{\dagger}(\bm{r})H_{e}^{\prime}\Psi(\bm{r}) \\
&= \frac{V_{e}^{\prime}}{2}\{[F_{n-1}(u) + F_{n}(u)]\mathrm{cos}(K\tilde{x}_{0}) \\
&\quad- 2s\xi\eta\rho_{0}R_{n}(u)\mathrm{sin}(K\tilde{x}_{0})\},
\end{split}
\end{equation}
with
\begin{equation}
\label{eq. 15}
\rho_{0} = \frac{C_{+}^{1/2}C_{-}^{1/2} - C_{+}^{-1/2}C_{-}^{-1/2}}{(C_{+} + C_{+}^{-1})^{1/2}(C_{-} + C_{-}^{-1})^{1/2}},
\end{equation}
\begin{equation}
\label{eq. 16}
F_{n}(u) = e^{-u/2}L_{n}(u),
\end{equation}
\begin{equation}
\label{eq. 17}
R_{n}(u) = \sqrt{\frac{2n}{u}}e^{-u/2}[L_{n-1}(u) - L_{n}(u)],
\end{equation}
where $L_{n}(u) = \sum\limits_{m = 0}^{n}\frac{(-1)^{m}n!u^{m}}{(m!)^{2}(n - m)!}$ is the Laguerre polynomial of order $n$ with $u = \theta\frac{K^{2}l^{2}}{2}$ and $\theta = \frac{v_{1}}{\sqrt{v_{2}^{2} + v_{+}^{2} - v_{-}^{2}}}$. Following Refs. \cite{32,33,41}, the diffusive conductivity can be simplified to an analytical form by using the higher Landau-level approximation
\begin{equation}
\label{eq. 18}
e^{-u/2}L_{n}(u) \rightarrow \frac{1}{\sqrt{\pi\sqrt{nu}}}\mathrm{cos}(2\sqrt{nu} - \frac{\pi}{4})
\end{equation}
and
\begin{equation}
\label{eq. 19}
R_{n}(u) \rightarrow \sqrt{2}\frac{1}{\sqrt{\pi\sqrt{nu}}}\mathrm{sin}(2\sqrt{nu} - \frac{\pi}{4}).
\end{equation}
The energy correction can be written as
\begin{equation}
\label{eq. 20}
\begin{split}
\Delta\epsilon_{e} = V_{e}^{\prime}\frac{1}{\sqrt{\pi\sqrt{nu}}}&[\mathrm{cos}(2\sqrt{nu} - \frac{\pi}{4})\mathrm{cos}(K\tilde{x}_{0}) \\
&- \sqrt{2}s\xi\eta\rho_{0}\mathrm{sin}(2\sqrt{nu} - \frac{\pi}{4})\mathrm{sin}(K\tilde{x}_{0})],
\end{split}
\end{equation}
which finally leads to the nonzero drift velocity
\begin{equation}
\label{eq. 21}
\begin{split}
v_{y} = \frac{1}{\hbar}\frac{\partial\Delta\epsilon_{e}}{\partial k_{y}} = -\frac{2V_{e}^{\prime}}{\hbar K}&u\frac{1}{\sqrt{\pi\sqrt{nu}}}[\mathrm{cos}(2\sqrt{nu} - \frac{\pi}{4})\mathrm{cos}(K\tilde{x}_{0}) \\
&- \sqrt{2}s\xi\eta\rho_{0}\mathrm{sin}(2\sqrt{nu} - \frac{\pi}{4})\mathrm{sin}(K\tilde{x}_{0})].
\end{split}
\end{equation}

The summation in Eq. (\ref{eq. 13}) can be represented by
\begin{equation}
\label{eq. 22}
\sum_{\zeta} = \frac{L_{y}}{2\pi}\int_{0}^{L_{x}/l^2}dk_{y}\sum_{s,\xi,\eta,n}.
\end{equation}
With the help of Eq. (\ref{eq. 21}), one can obtain the final expression for the diffusive dc conductivity
\begin{equation}
\label{eq. 23}
\sigma_{yy} = \frac{e^{2}}{h}W_{e}\Phi,
\end{equation}
where
\begin{equation}
\label{eq. 24}
W_{e} = \frac{V_{e}^{\prime 2}\tau_{0}\beta}{\hbar}
\end{equation}
is the dimensionless strength of the electric modulation and
\begin{equation}
\label{eq. 25}
\begin{split}
\Phi = \frac{\sqrt{u}}{\pi}\sum_{s,\xi,\eta,n}\frac{g}{(1 + g)^{2}}\frac{1}{\sqrt{n}}&[\mathrm{cos}^{2}(2\sqrt{nu} - \frac{\pi}{4}) \\
&+ 2\rho_{0}^{2}\mathrm{sin}^{2}(2\sqrt{nu} - \frac{\pi}{4})]
\end{split}
\end{equation}
is the dimensionless conductivity with the exponential function
\begin{equation}
\label{eq. 26}
g(\epsilon) = \mathrm{exp}[\beta(\epsilon - \epsilon_{F})].
\end{equation}
Here, we assume that the collisional time $\tau_{\epsilon_{\zeta}}$ is a constant $\tau_{0}$, which is a good approximation when temperature is sufficiently low \cite{32,33,40}.

\begin{figure}[h]
\centering
\subfigbottomskip=2pt
\subfigcapskip=-5pt
\subfigure[$E_{z} = 0$]{
\label{fig:1a}
\includegraphics[width=0.45\linewidth]{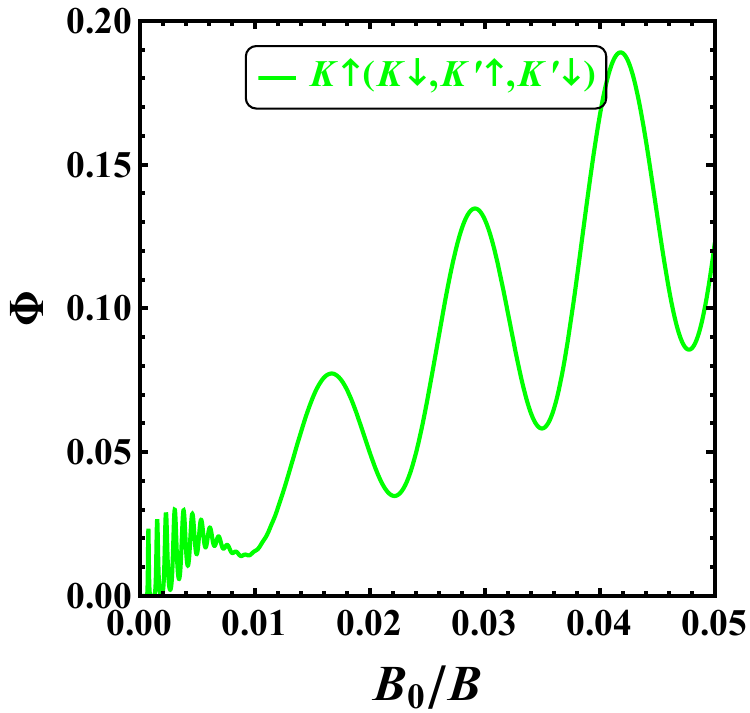}
}
\subfigure[$E_{z} = 0.2E_{c}$]{
\label{fig:1b}
\includegraphics[width=0.45\linewidth]{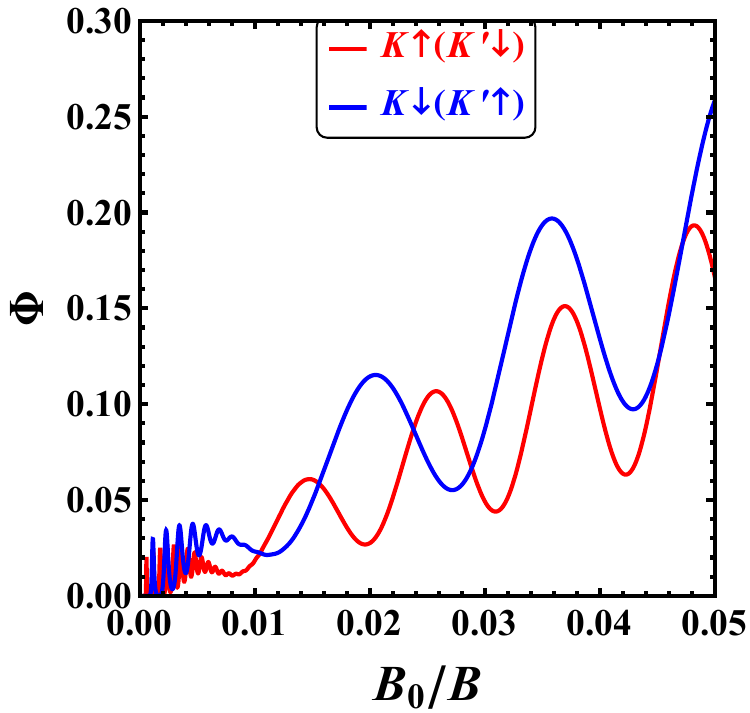}
}
 \\
\subfigure[$E_{z} = 0.4E_{c}$]{
\label{fig:1c}
\includegraphics[width=0.45\linewidth]{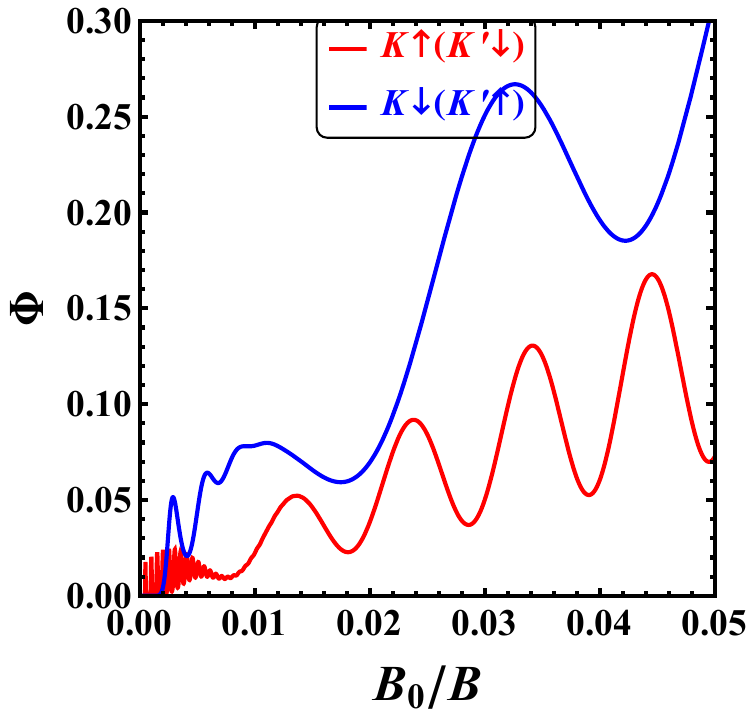}
}
\subfigure[$E_{z} = 0.2E_{c}$, $v_{-} = v_{+} = 0$]{
\label{fig:1d}
\includegraphics[width=0.45\linewidth]{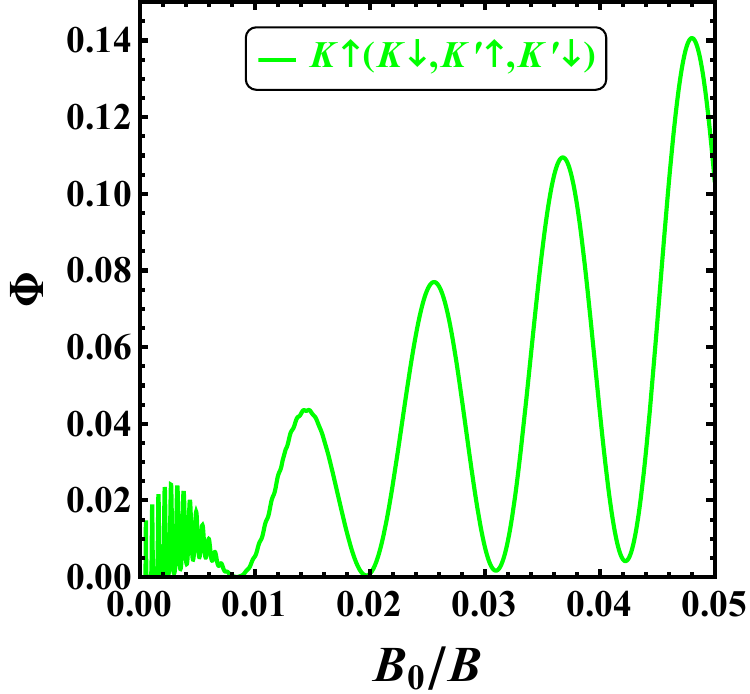}
}
\caption{\label{fig:1}The dimensionless conductivity versus the inverse magnetic field. The modulation period is $a_{0} = 350 \mathrm{nm}$, the temperature is $T = 3 \mathrm{K}$, and the Fermi energy is taken as 60 meV. $B_{0} = \frac{\hbar}{ea_{0}^{2}}$ is the characteristic magnetic field.}

\end{figure}

The dimensionless longitudinal diffusive magnetoconductivity $\Phi$ along the $y$ direction is calculated with Eq. (\ref{eq. 25}) and plotted in Figs.\ \ref{fig:1a}-\ref{fig:1c} as the function of the inverse magnetic field. It is shown in Fig.\ \ref{fig:1a} that the Weiss oscillation in monolayer 1{\it T}$^{\prime}$-\ce{MoS2} is spin-unpolarized and valley-unpolarized in the absence of the perpendicular electric field $E_{z}$. With $E_{z} = 0.2E_{c}$, as shown in Fig.\ \ref{fig:1b}, the curves split into two branches, resulting in the spin-valley polarization in the Weiss oscillation. The red curve in Fig.\ \ref{fig:1b} represents the Weiss oscillation for the spin-up electrons in the {\it K} valley or the spin-down electrons in the {\it K}$^{\prime}$ valley, while the blue curve represents that for the spin-down electrons in the {\it K} valley or the spin-up electrons in the {\it K}$^{\prime}$ valley. Comparing to Fig.\ \ref{fig:1a}, one finds that the period of the Weiss oscillation in the {\it K}$\uparrow$ and {\it K}$^{\prime}$$\downarrow$ channel decreases while that in the {\it K}$\downarrow$ and {\it K}$^{\prime}$$\uparrow$ channel increases with finite $E_{z}$. The result of $E_{z} = 0.4E_{c}$ is plotted in Fig.\ \ref{fig:1c}. One finds that the polarization is increased by imparting stronger perpendicular electric field $E_{z}$.

The spin-valley polarization in the Weiss oscillation can be well understood analytically. From Eq. (\ref{eq. 25}), one can find that the spin- and valley-dependences in the Weiss oscillation come from the Landau level $\epsilon_{n}$, which is determined by two polarization terms, $\xi\delta$ and $\delta^{2}$. These two terms represent the additional shift caused by the spin-orbit coupling and are proportional to the strength of the spin-orbit coupling gap. When the perpendicular electric field is zero ($\alpha = 0$), $\xi\delta$ and $\delta^{2}$ are no longer spin- or valley-dependent, leading to an unpolarized Weiss oscillation. When $\alpha$ is finite, both $\xi\delta$ and $\delta^{2}$ are dependent on $s\xi$, leading to a spin-valley-polarization behavior. Consequently, the spin-orbit coupling gap is one ingredient in the spin-valley polarization of the Weiss oscillation with a perpendicular electric field. Another ingredient is the tilted Dirac cones in monolayer 1{\it T}$^{\prime}$-\ce{MoS2}, which is represented by two velocity correction terms $v_{-}$ and $v_{+}$. One can easily find that when $v_{-}$ and $v_{+}$ vanish, the Landau level is unpolarized, resulting in an unpolarized Weiss oscillation. To be more clearly, we artificially take the parameters $E_{z} = 0.2E_{c}$ and $v_{-} = v_{+} = 0$, and plot the dimensionless conductivity in Fig.\ \ref{fig:1d}. It can be obviously seen that the Weiss oscillation is unpolarized now. Therefore the spin-valley polarization in the Weiss oscillation in monolayer 1{\it T}$^{\prime}$-\ce{MoS2} stems from the interplay between the tilted Dirac cones, the spin-orbit coupling gap, and the external electric potential modulation. In fact, silicene is a material with the spin-orbit coupling gap but is lack of the tilted Dirac cones and 8-{\it Pmmn} borophene is a material with the tilted Dirac cones but is lack of the spin-orbit coupling gap. Neither of them can exhibit the spin-valley polarization in the Weiss oscillation \cite{40,41} with a single electric potential modulation. Comparing to the spin- and valley-polarized Weiss oscillation in silicene \cite{40}, the tilted Dirac cones participate in the spin-valley-polarized Weiss oscillation in monolayer 1{\it T}$^{\prime}$-\ce{MoS2} instead of an additional spatial periodic electric field modulation along the $z$ direction.

\subsection{High Fermi level case}
\label{subsec:High Fermi level case}

Next, we consider the high Fermi level case, where the Fermi level is raised high enough so that the spin-orbit coupling gap can be neglected. Consequently, only the Landau levels with index $n$ high enough can contribute to the diffusive magnetoconductivity and Eq. (\ref{eq. 8}) is simplified as
\begin{equation}
\label{eq. 27}
\epsilon_{n} = \eta\sqrt{2n}\frac{\hbar v_{h}}{l}
\end{equation}
with
\begin{equation}
\label{eq. 28}
v_{h} = \sqrt{\frac{v_{1}(v_{2}^{2} + v_{+}^{2} - v_{-}^{2})^{3/2}}{v_{2}^{2} + v_{+}^{2}}},
\end{equation}
which is similar to that in pristine graphene \cite{32}.

\begin{figure}[h]
\centering
\includegraphics[width=0.9\linewidth]{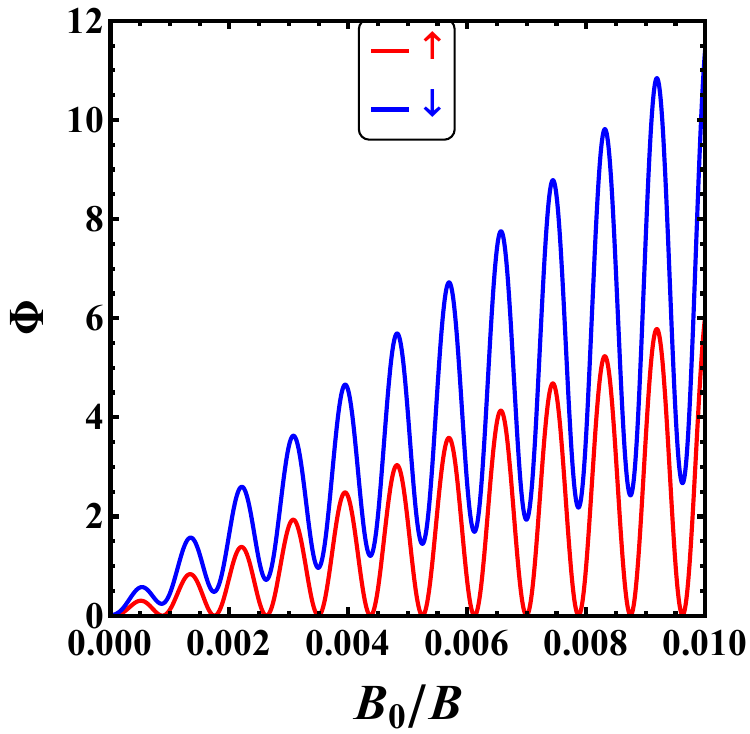}
\caption{\label{fig:2}The dimensionless conductivity versus the inverse magnetic field. The modulation period is $a_{0} = 350 \mathrm{nm}$, the temperature is $T = 3 \mathrm{K}$, and the Fermi energy is taken as 0.5 eV. $B_{0} = \frac{\hbar}{ea_{0}^{2}}$ is the characteristic magnetic field.}
\end{figure}

Now, in addition to the perpendicular magnetic field $B$, we introduce a magnetic modulation $ B^{\prime}\mathrm{cos}(Kx)\bm{\hat{e}_{z}}$ with $B^{\prime} \ll B$ and the modulation period along the $x$ direction $a_{0} = \frac{2\pi}{K}$. Under the Landau gauge, $\bm{A} = [0,Bx + \frac{B^{\prime}}{K}\mathrm{sin}(Kx),0]$, the perturbation Hamiltonian is given by
\begin{equation}
\label{eq. 29}	
H_{m}^{\prime} = \frac{eB^{\prime}\mathrm{sin}(Kx)}{K}[sv_{2}\sigma_{x} + \xi v_{-}\sigma_{0} + \xi v_{+}\sigma_{z}].
\end{equation}
The first-order energy correction is obtained as
\begin{equation}
\label{eq. 30}
\begin{split}
\Delta\epsilon_{m} &= \xi\frac{ev_{2}B^{\prime}}{K}\rho_{1}e^{-u/2}L_{n}(u)\mathrm{sin}(K\tilde{x}_{0}) \\
&- \eta\frac{ev_{2}B^{\prime}}{K}\rho_{2}R_{n}(u)\mathrm{cos}(K\tilde{x}_{0}),
\end{split}
\end{equation}
with
\begin{equation}
\label{eq. 31}
\begin{split}
\rho_{1} &= s(-\frac{1}{C_{+} + C_{+}^{-1}} + \frac{1}{C_{-} + C_{-}^{-1}}) + \frac{v_{-}}{v_{2}} \\
&+ s\frac{v_{+}}{v_{2}}\frac{1}{2}(\frac{C_{+} - C_{+}^{-1}}{C_{+} + C_{+}^{-1}} + \frac{C_{-} - C_{-}^{-1}}{C_{-} + C_{-}^{-1}}),
\end{split}
\end{equation}
\begin{equation}
\label{eq. 32}
\begin{split}
&\rho_{2} \\
&= \frac{C_{+}^{\frac{1}{2}}C_{-}^{-\frac{1}{2}} - C_{+}^{-\frac{1}{2}}C_{-}^{\frac{1}{2}}}{(C_{+} + C_{+}^{-1})^\frac{1}{2}(C_{-} + C_{-}^{-1})^\frac{1}{2}} + s\frac{v_{-}}{v_{2}}\frac{C_{+}^{\frac{1}{2}}C_{-}^{\frac{1}{2}} - C_{+}^{-\frac{1}{2}}C_{-}^{-\frac{1}{2}}}{(C_{+} + C_{+}^{-1})^\frac{1}{2}(C_{-} + C_{-}^{-1})^\frac{1}{2}} \\
&+ \frac{v_{+}}{v_{2}}\frac{C_{+}^{\frac{1}{2}}C_{-}^{\frac{1}{2}} + C_{+}^{-\frac{1}{2}}C_{-}^{-\frac{1}{2}}}{(C_{+} + C_{+}^{-1})^\frac{1}{2}(C_{-} + C_{-}^{-1})^\frac{1}{2}}.
\end{split}
\end{equation}
We note that $\rho_{1,2}$ and therefore $\Delta\epsilon_{m}$ are spin-dependent, which are different from those results in 8-{\it Pmmn} borophene \cite{41}. Following the approaches in Refs. \cite{32,41}, the longitudinal diffusive magnetoconductivity can be obtained as
\begin{equation}
\label{eq. 33}
\sigma_{yy} = \frac{e^{2}}{h}W_{m}\Phi,
\end{equation}
where
\begin{equation}
\label{eq. 34}
W_{m} = (\frac{ev_{2}B^{\prime}}{K})^{2}\frac{\tau_{0}\beta}{\hbar}
\end{equation}
is the dimensionless strength of the magnetic modulation and
\begin{equation}
\label{eq. 35}
\begin{split}
\Phi &= \frac{T}{4\pi^{2}T_{D}}[(\rho_{1}^{2} + 2\rho_{2}^{2}) \\
&+ (\rho_{1}^{2} - 2\rho_{2}^{2})\mathrm{cos}(\beta\epsilon_{F}\frac{T}{T_{D}} - \frac{\pi}{2})\frac{T/T_{D}}{\mathrm{sinh}(T/T_{D})}]
\end{split}
\end{equation}
is the dimensionless conductivity with $T_{D} = \frac{\hbar v_{F}B}{4 \pi^{2}a_{0}k_{B}B_{0}}$.

We plot $\Phi$ as the function of the inverse magnetic field in Fig.\ \ref{fig:2}. It is shown in Fig.\ \ref{fig:2} that the Weiss oscillation in monolayer 1{\it T}$^{\prime}$-\ce{MoS2} is spin-polarized. The conductivities for the spin-up and spin-down electrons oscillate with the same period but different amplitudes.

The spin-polarization of the Weiss oscillation in Fig.\ \ref{fig:2} comes from the spin-dependence of $\rho_{1}^{2}$ and $\rho_{2}^{2}$. One can easily find that, without a tilted Dirac cone, $v_{-} = 0$, and $\rho_{1}^{2}$, $\rho_{2}^{2}$ are spin-independent, resulting in an spin-unpolarized $\Phi$ in Eq. (\ref{eq. 35}). However, merely the presence of the tilted Dirac cones can not ensure the spin-polarization of the Weiss oscillation. Another key ingredient is the spin-dependence of the low-energy effective Hamiltonian (Eq. (\ref{eq. 5})), which comes from the spin-orbit coupling and leads to the spin-dependence of $\rho_{1}$ and $\rho_{2}$, even without a finite spin-orbit coupling gap. Unlike silicene \cite{40}, where the spin-orbit coupling participate in the spin-polarized Weiss oscillation merely in the form of a finite spin-orbit coupling gap, the spin-polarized Weiss oscillation in monolayer 1{\it T}$^{\prime}$-\ce{MoS2} arises from the interplay between the tilted Dirac cones, the spin-orbit coupling, and the external magnetic filed modulation, even though the spin-orbit coupling gap can be neglected compared to the Landau levels near the high Fermi level.

\section{\label{sec:level4}Summary}
\label{sec:4}

In this work, we have theoretically studied the spin and valley polarization in the Weiss oscillations of the monolayer 1{\it T}$^{\prime}$-\ce{MoS2} and find two unique behaviors.

Firstly, when the Fermi level is close to the spin-orbit coupling gap, in the presence of both a spatial periodic electric potential modulation in the lateral direction and a nonzero perpendicular electric field, the Weiss oscillation in monolayer 1{\it T}$^{\prime}$-\ce{MoS2} splits into two branches and exhibits spin-valley polarization. The spin-valley polarization stems from the interplay between the tilted Dirac cones, the spin-orbit coupling gap, and the external electric potential modulation, and can be treated as a signature of monolayer 1{\it T}$^{\prime}$-\ce{MoS2}.

Secondly, when the Fermi level is high enough and the spin-orbit coupling gap can be neglected, the Weiss oscillation is spin-polarized with a magnetic field modulation in the lateral direction, which stems from the interplay between the tilted Dirac cones, the spin-orbit coupling, and the external magnetic field modulation.

\begin{acknowledgments}
This work is supported by the National Key R\&D Program of China (Grant No. 2022YFA1403601).
\end{acknowledgments}

\begin{widetext}

\begin{appendix}

\section{ALGEBRAIC SOLUTION OF THE LANDAU LEVELS}
\label{app:ALGEBRAIC SOLUTION OF THE LANDAU LEVELS}

In order to find the Landau levels and the eigenfunctions, we follow the algebraic approach in Ref. \cite{42}. Taking the spin-up ($s = +1$) scenario as an example, Eq. (\ref{eq. 7}) in the text can be rewritten as
\begin{equation}
\label{eq. A1}
H = \begin{pmatrix}
-\xi(E_{+} + E_{-})(a + a^{\dagger} + \delta) & -E_{B}a \\
-E_{B}a^{\dagger} & \xi(E_{+} - E_{-})(a + a^{\dagger} +\delta)
\end{pmatrix},
\end{equation}with $a = \frac{1}{\sqrt{2}}(X + \partial_{X})$, $a^{\dagger} = \frac{1}{\sqrt{2}}(X - \partial_{X})$, $E_{B} = \frac{\sqrt{2}\hbar v_{c}}{l}$, $E_{-} = \frac{\hbar v_{-}}{\sqrt{2}l}\sqrt{\frac{v_{1}}{v_{2}}}$, $E_{+} = \frac{\hbar v_{+}}{\sqrt{2}l}\sqrt{\frac{v_{1}}{v_{2}}}$, and $\delta = \sqrt{2}(s\alpha - \xi)\frac{\Delta_{so}l}{\hbar v_{2}}$.

We define an operator
\begin{equation}
\label{eq. A2}
\bar{H} = C\sigma_{x}\sigma_{z}H\sigma_{z}\sigma_{x}C = \begin{pmatrix}
\xi(E_{+} - E_{-})(a + a^{\dagger} + \delta) & E_{B}a \\
E_{B}a^{\dagger} & -\xi(E_{+} + E_{-})(a + a^{\dagger} +\delta)
\end{pmatrix},
\end{equation}
where $C$ stands for the complex conjugate. 
One can easily verify that the solution of the eigenproblem $H\psi = \epsilon \psi$ satisfies
\begin{equation}
\label{eq. A3}
[\epsilon (H + \bar{H}) - \bar{H} H] \psi = \epsilon^{2} \psi,
\end{equation}
which can be rewritten as
\begin{equation}
\label{eq. A4}
(J \sigma_{0} + \tilde{K}) \psi = \epsilon^{2} \psi,
\end{equation}with
\begin{equation}
\label{eq. A5}
J = [2(E_{+}^{2} - E_{-}^{2}) + E_{B}^{2}]aa^{\dagger} + (E_{+}^{2} - E_{-}^{2})(aa + a^{\dagger}a^{\dagger}) - 2[\xi\epsilon E_{-} - \delta(E_{+}^{2} - E_{-}^{2})](a + a^{\dagger})
\end{equation} and
\begin{equation}
\label{eq. A6}
\tilde{K} = \begin{pmatrix}
-2\xi\delta\epsilon E_{-} + (1 + \delta^{2})(E_{+}^{2} - E_{-}^{2}) + E_{B}^{2} & -\xi(E_{+} - E_{-})E_{B} \\
-\xi(E_{+} + E_{-})E_{B} & -2\xi\delta\epsilon E_{-} + (1 + \delta^{2})(E_{+}^{2} - E_{-}^{2})
\end{pmatrix}.
\end{equation}

To solve Eq. (\ref{eq. A4}), we take the ansatz $\psi = \phi\begin{pmatrix}u \\v\end{pmatrix}$, which yields 
\begin{equation}
\label{eq. A7}
(\tilde{K} - \lambda)\begin{pmatrix}
u \\
v
\end{pmatrix} = 0
\end{equation}
and
\begin{equation}
\label{eq. A8}
J\phi = (\epsilon^{2} - \lambda)\phi.
\end{equation}
The solution of Eq. (\ref{eq. A7}) reads
\begin{equation}
\label{eq. A9}
\lambda_{\pm} = -2\xi\delta E_{-}\epsilon + (1 + \delta^{2})(E_{+}^{2} - E_{-}^{2}) + \frac{1}{2}[E_{B}^{2} \pm E_{B}\sqrt{E_{B}^{2} + 4(E_{+}^2 - E_{-}^{2})}],
\end{equation}
\begin{equation}
\label{eq. A10}
\frac{u}{v} = \mp\xi C_{\pm}, C_{\pm} = \frac{2(E_{+} - E_{-})}{\sqrt{E_{B}^{2} + 4(E_{+}^{2} - E_{-}^{2})} \mp E_{B}}.
\end{equation}
By defining
\begin{equation}
\label{eq. A11}
\gamma = a \cosh{U} + a^{\dagger} \sinh{U} + \frac{C_{1}}{\omega}
\end{equation}
with
\begin{equation}
\label{eq. A12}
(\cosh{U})^{2} = \frac{1}{2}[1 + (2 E_{+}^{2} - 2 E_{-}^{2} + E_{B}^{2})/\omega],
\end{equation}
\begin{equation}
\label{eq. A13}
(\sinh{U})^{2} = \frac{1}{2}[(2 E_{+}^{2} - 2 E_{-}^{2} + E_{B}^{2})/\omega - 1],
\end{equation}
\begin{equation}
\label{eq. A14}
\omega = E_{B}\sqrt{4(E_{+}^{2} - E_{-}^{2}) + E_{B}^{2}},
\end{equation}
\begin{equation}
\label{eq. A15}
C_{1} = -2[\xi E_{-} \epsilon - \delta(E_{+}^{2} - E_{-}^{2})](\cosh{U} - \sinh{U}),
\end{equation}
the operator $J$ can be diagonalized as
\begin{equation}
\label{eq. A16}
J = \omega\gamma^{\dagger}\gamma -\frac{C_{1}^{2}}{\omega} + C_{2}
\end{equation}
with
\begin{equation}
\label{eq. A17}
C_{2} = \frac{1}{2} [\omega - 2(E_{+}^{2} - E_{-}^{2}) - E_{B}^{2}].
\end{equation}
Then, the eigenvalue of the operator $J$ can be obtained as
\begin{equation}
\label{eq. A18}
\omega_{n} = n \omega -\frac{C_{1}^{2}}{\omega} + C_{2}.
\end{equation}
In the coordinate representation, the annihilation operator reads $\gamma = \frac{1}{\sqrt{2}}(\tilde{X} + \partial_{\tilde{X}})$ with
\begin{equation}
\label{eq. A19}
\begin{split}
\tilde{X} &= (\cosh{U} + \sinh{U})X + \frac{\sqrt{2}C_{1}}{\omega} \\
&= \frac{1}{l}(\frac{\sqrt{v_{2}^{2} + v_{+}^{2} - v_{-}^{2}}}{v_{1}})^{1/2}(x + k_{y}l^{2} - \frac{\delta l}{\sqrt{2}} - \sqrt{2}\frac{\xi v_{-}\sqrt{\frac{v_{2}}{v_{1}}}\frac{\epsilon_{n}l^{2}}{\sqrt{2}\hbar} - \frac{\delta l}{2}(v_{+}^{2} - v_{-}^{2})}{v_{2}^{2} + v_{+}^{2} -v_{-}^{2}}).
\end{split}
\end{equation}
Therefore, the eigenfunction of $J$ is $\phi_{n}(\tilde{X})$ with $\phi_{n}$ being the harmonic oscillator wave function.

From Eq. (\ref{eq. A8}, \ref{eq. A9}, \ref{eq. A18}) and Eq. (\ref{eq. A10}), we obtain the eigenvalue
\begin{equation}
\label{eq. A20}
\epsilon = \frac{\sqrt{2}\hbar v_{c}}{l}\frac{1}{v_{2}^{2} + v_{+}^{2}}[-\frac{1}{2}\xi\delta v_{2}v_{-} + \eta\sqrt{n\frac{(v_{2}^{2} + v_{+}^{2})(v_{2}^{2} + v_{+}^{2} -v_{-}^{2})^{3/2}}{v_{2}} + \frac{\delta^{2}}{4}v_{+}^{2}(v_{2}^{2} + v_{+}^{2} - v_{-}^{2})}],
\end{equation}
and the eigenfunction
\begin{equation}
\label{eq. A21}
\psi = \mu \chi_{1} + \nu \chi_{2}
\end{equation}
being the superposition of the two basis vectors
\begin{equation}
\label{eq. A22}
\chi_{1} = \frac{1}{\sqrt{2}}\phi_{n-1}\begin{pmatrix}
-\xi(\frac{C_{+}}{C_{+} + C_{+}^{-1}})^{1/2} \\
(\frac{C_{+}^{-1}}{C_{+} + C_{+}^{-1}})^{1/2}
\end{pmatrix},\quad   
\chi_{2} = \frac{1}{\sqrt{2}}\phi_{n}\begin{pmatrix}
(\frac{C_{-}}{C_{-} + C_{-}^{-1}})^{1/2} \\
\xi(\frac{C_{-}^{-1}}{C_{-} + C_{-}^{-1}})^{1/2}
\end{pmatrix}.
\end{equation}
In order to determine the coefficients $\mu$ and $\nu$, we consider the asymptotic situation of $v_{-} = v_{+} = 0$. In this case, the wave function is similar to that of pristine graphene \cite{32} and is given by
\begin{equation}
\label{eq. A23}
\psi(v_{-} \rightarrow 0, v_{+} \rightarrow 0) = \frac{1}{\sqrt{2}}\begin{pmatrix}
-\phi_{n-1}(X) \\
\eta\phi_{n}(X)
\end{pmatrix},
\end{equation}
where $\eta = +1(-1)$ labels the band index.
Comparing Eq. (\ref{eq. A21}) under the limit $v_{-} = v_{+} = 0$ with Eq. (\ref{eq. A23}), we find $\mu = 1$ and $\nu = \eta$, leading to the eigenfunction
\begin{equation}
\label{eq. A24}
\psi = \frac{1}{\sqrt{2}}\begin{pmatrix}
-\xi(\frac{C_{+}}{C_{+} + C_{+}^{-1}})^{1/2}\phi_{n-1} + \eta(\frac{C_{-}}{C_{-} + C_{-}^{-1}})^{1/2}\phi_{n} \\
(\frac{C_{+}^{-1}}{C_{+} + C_{+}^{-1}})^{1/2}\phi_{n-1} + \eta\xi(\frac{C_{-}^{-1}}{C_{-} + C_{-}^{-1}})^{1/2}\phi_{n}
\end{pmatrix}.
\end{equation}

\end{appendix}

\end{widetext}

\bibliography{Weiss_1T_-MoS2.bib}

\begin{thebibliography}{45}%
\makeatletter
\providecommand \@ifxundefined [1]{%
 \@ifx{#1\undefined}
}%
\providecommand \@ifnum [1]{%
 \ifnum #1\expandafter \@firstoftwo
 \else \expandafter \@secondoftwo
 \fi
}%
\providecommand \@ifx [1]{%
 \ifx #1\expandafter \@firstoftwo
 \else \expandafter \@secondoftwo
 \fi
}%
\providecommand \natexlab [1]{#1}%
\providecommand \enquote  [1]{``#1''}%
\providecommand \bibnamefont  [1]{#1}%
\providecommand \bibfnamefont [1]{#1}%
\providecommand \citenamefont [1]{#1}%
\providecommand \href@noop [0]{\@secondoftwo}%
\providecommand \href [0]{\begingroup \@sanitize@url \@href}%
\providecommand \@href[1]{\@@startlink{#1}\@@href}%
\providecommand \@@href[1]{\endgroup#1\@@endlink}%
\providecommand \@sanitize@url [0]{\catcode `\\12\catcode `\$12\catcode
  `\&12\catcode `\#12\catcode `\^12\catcode `\_12\catcode `\%12\relax}%
\providecommand \@@startlink[1]{}%
\providecommand \@@endlink[0]{}%
\providecommand \url  [0]{\begingroup\@sanitize@url \@url }%
\providecommand \@url [1]{\endgroup\@href {#1}{\urlprefix }}%
\providecommand \urlprefix  [0]{URL }%
\providecommand \Eprint [0]{\href }%
\providecommand \doibase [0]{https://doi.org/}%
\providecommand \selectlanguage [0]{\@gobble}%
\providecommand \bibinfo  [0]{\@secondoftwo}%
\providecommand \bibfield  [0]{\@secondoftwo}%
\providecommand \translation [1]{[#1]}%
\providecommand \BibitemOpen [0]{}%
\providecommand \bibitemStop [0]{}%
\providecommand \bibitemNoStop [0]{.\EOS\space}%
\providecommand \EOS [0]{\spacefactor3000\relax}%
\providecommand \BibitemShut  [1]{\csname bibitem#1\endcsname}%
\let\auto@bib@innerbib\@empty
\bibitem [{\citenamefont {Qian}\ \emph {et~al.}(2014)\citenamefont {Qian},
  \citenamefont {Liu}, \citenamefont {Fu},\ and\ \citenamefont {Li}}]{1}%
  \BibitemOpen
  \bibfield  {author} {\bibinfo {author} {\bibfnamefont {X.}~\bibnamefont
  {Qian}}, \bibinfo {author} {\bibfnamefont {J.}~\bibnamefont {Liu}}, \bibinfo
  {author} {\bibfnamefont {L.}~\bibnamefont {Fu}},\ and\ \bibinfo {author}
  {\bibfnamefont {J.}~\bibnamefont {Li}},\ }\href
  {https://doi.org/10.1126/science.1256815} {\bibfield  {journal} {\bibinfo
  {journal} {Science}\ }\textbf {\bibinfo {volume} {346}},\ \bibinfo {pages}
  {1344} (\bibinfo {year} {2014})}\BibitemShut {NoStop}%
\bibitem [{\citenamefont {Xiao}\ \emph {et~al.}(2012)\citenamefont {Xiao},
  \citenamefont {Liu}, \citenamefont {Feng}, \citenamefont {Xu},\ and\
  \citenamefont {Yao}}]{2}%
  \BibitemOpen
  \bibfield  {author} {\bibinfo {author} {\bibfnamefont {D.}~\bibnamefont
  {Xiao}}, \bibinfo {author} {\bibfnamefont {G.-B.}\ \bibnamefont {Liu}},
  \bibinfo {author} {\bibfnamefont {W.}~\bibnamefont {Feng}}, \bibinfo {author}
  {\bibfnamefont {X.}~\bibnamefont {Xu}},\ and\ \bibinfo {author}
  {\bibfnamefont {W.}~\bibnamefont {Yao}},\ }\href
  {https://doi.org/10.1103/PhysRevLett.108.196802} {\bibfield  {journal}
  {\bibinfo  {journal} {Phys. Rev. Lett.}\ }\textbf {\bibinfo {volume} {108}},\
  \bibinfo {pages} {196802} (\bibinfo {year} {2012})}\BibitemShut {NoStop}%
\bibitem [{\citenamefont {Wang}\ \emph {et~al.}(2012)\citenamefont {Wang},
  \citenamefont {Kalantar-Zadeh}, \citenamefont {Kis}, \citenamefont
  {Coleman},\ and\ \citenamefont {Strano}}]{3}%
  \BibitemOpen
  \bibfield  {author} {\bibinfo {author} {\bibfnamefont {Q.~H.}\ \bibnamefont
  {Wang}}, \bibinfo {author} {\bibfnamefont {K.}~\bibnamefont
  {Kalantar-Zadeh}}, \bibinfo {author} {\bibfnamefont {A.}~\bibnamefont {Kis}},
  \bibinfo {author} {\bibfnamefont {J.~N.}\ \bibnamefont {Coleman}},\ and\
  \bibinfo {author} {\bibfnamefont {M.~S.}\ \bibnamefont {Strano}},\ }\href
  {https://doi.org/10.1038/nnano.2012.193} {\bibfield  {journal} {\bibinfo
  {journal} {Nature Nanotechnology}\ ,\ \bibinfo {pages} {699}} (\bibinfo
  {year} {2012})}\BibitemShut {NoStop}%
\bibitem [{\citenamefont {Xu}\ \emph {et~al.}(2014)\citenamefont {Xu},
  \citenamefont {Yao}, \citenamefont {Xiao},\ and\ \citenamefont {Heinz}}]{4}%
  \BibitemOpen
  \bibfield  {author} {\bibinfo {author} {\bibfnamefont {X.}~\bibnamefont
  {Xu}}, \bibinfo {author} {\bibfnamefont {W.}~\bibnamefont {Yao}}, \bibinfo
  {author} {\bibfnamefont {D.}~\bibnamefont {Xiao}},\ and\ \bibinfo {author}
  {\bibfnamefont {T.~F.}\ \bibnamefont {Heinz}},\ }\href
  {https://doi.org/10.1038/nphys2942} {\bibfield  {journal} {\bibinfo
  {journal} {Nature Physics}\ }\textbf {\bibinfo {volume} {10}},\ \bibinfo
  {pages} {343} (\bibinfo {year} {2014})}\BibitemShut {NoStop}%
\bibitem [{\citenamefont {Manzeli}\ \emph {et~al.}(2017)\citenamefont
  {Manzeli}, \citenamefont {Ovchinnikov}, \citenamefont {Pasquier},
  \citenamefont {Yazyev},\ and\ \citenamefont {Kis}}]{5}%
  \BibitemOpen
  \bibfield  {author} {\bibinfo {author} {\bibfnamefont {S.}~\bibnamefont
  {Manzeli}}, \bibinfo {author} {\bibfnamefont {D.}~\bibnamefont
  {Ovchinnikov}}, \bibinfo {author} {\bibfnamefont {D.}~\bibnamefont
  {Pasquier}}, \bibinfo {author} {\bibfnamefont {O.~V.}\ \bibnamefont
  {Yazyev}},\ and\ \bibinfo {author} {\bibfnamefont {A.}~\bibnamefont {Kis}},\
  }\href {https://doi.org/10.1038/natrevmats.2017.33} {\bibfield  {journal}
  {\bibinfo  {journal} {Nature Reviews Materials}\ }\textbf {\bibinfo {volume}
  {2}},\ \bibinfo {pages} {17033} (\bibinfo {year} {2017})}\BibitemShut
  {NoStop}%
\bibitem [{\citenamefont {Wu}\ \emph {et~al.}(2018)\citenamefont {Wu},
  \citenamefont {Fatemi}, \citenamefont {Gibson}, \citenamefont {Watanabe},
  \citenamefont {Taniguchi}, \citenamefont {Cava},\ and\ \citenamefont
  {Jarillo-Herrero}}]{6}%
  \BibitemOpen
  \bibfield  {author} {\bibinfo {author} {\bibfnamefont {S.}~\bibnamefont
  {Wu}}, \bibinfo {author} {\bibfnamefont {V.}~\bibnamefont {Fatemi}}, \bibinfo
  {author} {\bibfnamefont {Q.~D.}\ \bibnamefont {Gibson}}, \bibinfo {author}
  {\bibfnamefont {K.}~\bibnamefont {Watanabe}}, \bibinfo {author}
  {\bibfnamefont {T.}~\bibnamefont {Taniguchi}}, \bibinfo {author}
  {\bibfnamefont {R.~J.}\ \bibnamefont {Cava}},\ and\ \bibinfo {author}
  {\bibfnamefont {P.}~\bibnamefont {Jarillo-Herrero}},\ }\href
  {https://doi.org/10.1126/science.aan6003} {\bibfield  {journal} {\bibinfo
  {journal} {Science}\ }\textbf {\bibinfo {volume} {359}},\ \bibinfo {pages}
  {76} (\bibinfo {year} {2018})}\BibitemShut {NoStop}%
\bibitem [{\citenamefont {Sajadi}\ \emph {et~al.}(2018)\citenamefont {Sajadi},
  \citenamefont {Palomaki}, \citenamefont {Fei}, \citenamefont {Zhao},
  \citenamefont {Bement}, \citenamefont {Olsen}, \citenamefont {Luescher},
  \citenamefont {Xu}, \citenamefont {Folk},\ and\ \citenamefont {Cobden}}]{7}%
  \BibitemOpen
  \bibfield  {author} {\bibinfo {author} {\bibfnamefont {E.}~\bibnamefont
  {Sajadi}}, \bibinfo {author} {\bibfnamefont {T.}~\bibnamefont {Palomaki}},
  \bibinfo {author} {\bibfnamefont {Z.}~\bibnamefont {Fei}}, \bibinfo {author}
  {\bibfnamefont {W.}~\bibnamefont {Zhao}}, \bibinfo {author} {\bibfnamefont
  {P.}~\bibnamefont {Bement}}, \bibinfo {author} {\bibfnamefont
  {C.}~\bibnamefont {Olsen}}, \bibinfo {author} {\bibfnamefont
  {S.}~\bibnamefont {Luescher}}, \bibinfo {author} {\bibfnamefont
  {X.}~\bibnamefont {Xu}}, \bibinfo {author} {\bibfnamefont {J.~A.}\
  \bibnamefont {Folk}},\ and\ \bibinfo {author} {\bibfnamefont {D.~H.}\
  \bibnamefont {Cobden}},\ }\href {https://doi.org/10.1126/science.aaar4426}
  {\bibfield  {journal} {\bibinfo  {journal} {Science}\ }\textbf {\bibinfo
  {volume} {362}},\ \bibinfo {pages} {922} (\bibinfo {year}
  {2018})}\BibitemShut {NoStop}%
\bibitem [{\citenamefont {Ominato}\ \emph {et~al.}(2020)\citenamefont
  {Ominato}, \citenamefont {Fujimoto},\ and\ \citenamefont {Matsuo}}]{8}%
  \BibitemOpen
  \bibfield  {author} {\bibinfo {author} {\bibfnamefont {Y.}~\bibnamefont
  {Ominato}}, \bibinfo {author} {\bibfnamefont {J.}~\bibnamefont {Fujimoto}},\
  and\ \bibinfo {author} {\bibfnamefont {M.}~\bibnamefont {Matsuo}},\ }\href
  {https://doi.org/10.1103/PhysRevLett.124.166803} {\bibfield  {journal}
  {\bibinfo  {journal} {Phys. Rev. Lett.}\ }\textbf {\bibinfo {volume} {124}},\
  \bibinfo {pages} {166803} (\bibinfo {year} {2020})}\BibitemShut {NoStop}%
\bibitem [{\citenamefont {Tang}\ \emph {et~al.}(2017)\citenamefont {Tang},
  \citenamefont {Zhang}, \citenamefont {Wong}, \citenamefont {Pedramrazi},
  \citenamefont {Tsai}, \citenamefont {Jia}, \citenamefont {Moritz},
  \citenamefont {Claassen}, \citenamefont {Ryu}, \citenamefont {Kahn},
  \citenamefont {Jiang}, \citenamefont {Yan}, \citenamefont {Hashimoto},
  \citenamefont {Lu}, \citenamefont {Hwang}, \citenamefont {Hwang},
  \citenamefont {Hussain}, \citenamefont {Ugeda}, \citenamefont {Liu},
  \citenamefont {Xie}, \citenamefont {Devereaux}, \citenamefont {Crommie},
  \citenamefont {Mo},\ and\ \citenamefont {Shen}}]{9}%
  \BibitemOpen
  \bibfield  {author} {\bibinfo {author} {\bibfnamefont {S.}~\bibnamefont
  {Tang}}, \bibinfo {author} {\bibfnamefont {C.}~\bibnamefont {Zhang}},
  \bibinfo {author} {\bibfnamefont {D.}~\bibnamefont {Wong}}, \bibinfo {author}
  {\bibfnamefont {Z.}~\bibnamefont {Pedramrazi}}, \bibinfo {author}
  {\bibfnamefont {H.-Z.}\ \bibnamefont {Tsai}}, \bibinfo {author}
  {\bibfnamefont {C.}~\bibnamefont {Jia}}, \bibinfo {author} {\bibfnamefont
  {B.}~\bibnamefont {Moritz}}, \bibinfo {author} {\bibfnamefont
  {M.}~\bibnamefont {Claassen}}, \bibinfo {author} {\bibfnamefont
  {H.}~\bibnamefont {Ryu}}, \bibinfo {author} {\bibfnamefont {S.}~\bibnamefont
  {Kahn}}, \bibinfo {author} {\bibfnamefont {J.}~\bibnamefont {Jiang}},
  \bibinfo {author} {\bibfnamefont {H.}~\bibnamefont {Yan}}, \bibinfo {author}
  {\bibfnamefont {M.}~\bibnamefont {Hashimoto}}, \bibinfo {author}
  {\bibfnamefont {R.~G.}\ \bibnamefont {Lu}, \bibfnamefont {Donghui~Moore}},
  \bibinfo {author} {\bibfnamefont {C.-C.}\ \bibnamefont {Hwang}}, \bibinfo
  {author} {\bibfnamefont {C.}~\bibnamefont {Hwang}}, \bibinfo {author}
  {\bibfnamefont {Y.}~\bibnamefont {Hussain}, \bibfnamefont {Zahid~andChen}},
  \bibinfo {author} {\bibfnamefont {M.~M.}\ \bibnamefont {Ugeda}}, \bibinfo
  {author} {\bibfnamefont {Z.}~\bibnamefont {Liu}}, \bibinfo {author}
  {\bibfnamefont {X.}~\bibnamefont {Xie}}, \bibinfo {author} {\bibfnamefont
  {T.~P.}\ \bibnamefont {Devereaux}}, \bibinfo {author} {\bibfnamefont {M.~F.}\
  \bibnamefont {Crommie}}, \bibinfo {author} {\bibfnamefont {S.-K.}\
  \bibnamefont {Mo}},\ and\ \bibinfo {author} {\bibfnamefont {Z.-X.}\
  \bibnamefont {Shen}},\ }\href {https://doi.org/10.1038/nphys4174} {\bibfield
  {journal} {\bibinfo  {journal} {Nature Physics}\ }\textbf {\bibinfo {volume}
  {13}},\ \bibinfo {pages} {683} (\bibinfo {year} {2017})}\BibitemShut
  {NoStop}%
\bibitem [{\citenamefont {Jia}\ \emph {et~al.}(2017)\citenamefont {Jia},
  \citenamefont {Song}, \citenamefont {Li}, \citenamefont {Ran}, \citenamefont
  {Lu}, \citenamefont {Zheng}, \citenamefont {Zhu}, \citenamefont {Shi},
  \citenamefont {Sun}, \citenamefont {Wen}, \citenamefont {Xing},\ and\
  \citenamefont {Li}}]{10}%
  \BibitemOpen
  \bibfield  {author} {\bibinfo {author} {\bibfnamefont {Z.-Y.}\ \bibnamefont
  {Jia}}, \bibinfo {author} {\bibfnamefont {Y.-H.}\ \bibnamefont {Song}},
  \bibinfo {author} {\bibfnamefont {X.-B.}\ \bibnamefont {Li}}, \bibinfo
  {author} {\bibfnamefont {K.}~\bibnamefont {Ran}}, \bibinfo {author}
  {\bibfnamefont {P.}~\bibnamefont {Lu}}, \bibinfo {author} {\bibfnamefont
  {H.-J.}\ \bibnamefont {Zheng}}, \bibinfo {author} {\bibfnamefont {X.-Y.}\
  \bibnamefont {Zhu}}, \bibinfo {author} {\bibfnamefont {Z.-Q.}\ \bibnamefont
  {Shi}}, \bibinfo {author} {\bibfnamefont {J.}~\bibnamefont {Sun}}, \bibinfo
  {author} {\bibfnamefont {J.}~\bibnamefont {Wen}}, \bibinfo {author}
  {\bibfnamefont {D.}~\bibnamefont {Xing}},\ and\ \bibinfo {author}
  {\bibfnamefont {S.-C.}\ \bibnamefont {Li}},\ }\href
  {https://doi.org/10.1103/PhysRevB.96.041108} {\bibfield  {journal} {\bibinfo
  {journal} {Phys. Rev. B}\ }\textbf {\bibinfo {volume} {96}},\ \bibinfo
  {pages} {041108} (\bibinfo {year} {2017})}\BibitemShut {NoStop}%
\bibitem [{\citenamefont {Peng}\ \emph {et~al.}(2017)\citenamefont {Peng},
  \citenamefont {Yuan}, \citenamefont {Li}, \citenamefont {Yang}, \citenamefont
  {Xian}, \citenamefont {Yi}, \citenamefont {Shi},\ and\ \citenamefont
  {Fu}}]{11}%
  \BibitemOpen
  \bibfield  {author} {\bibinfo {author} {\bibfnamefont {L.}~\bibnamefont
  {Peng}}, \bibinfo {author} {\bibfnamefont {Y.}~\bibnamefont {Yuan}}, \bibinfo
  {author} {\bibfnamefont {G.}~\bibnamefont {Li}}, \bibinfo {author}
  {\bibfnamefont {X.}~\bibnamefont {Yang}}, \bibinfo {author} {\bibfnamefont
  {J.-J.}\ \bibnamefont {Xian}}, \bibinfo {author} {\bibfnamefont {C.-J.}\
  \bibnamefont {Yi}}, \bibinfo {author} {\bibfnamefont {Y.-G.}\ \bibnamefont
  {Shi}},\ and\ \bibinfo {author} {\bibfnamefont {Y.-S.}\ \bibnamefont {Fu}},\
  }\href {https://doi.org/10.1038/s41467-017-00745-8} {\bibfield  {journal}
  {\bibinfo  {journal} {Nature Communications}\ }\textbf {\bibinfo {volume}
  {8}},\ \bibinfo {pages} {659} (\bibinfo {year} {2017})}\BibitemShut {NoStop}%
\bibitem [{\citenamefont {Das}\ \emph {et~al.}(2020)\citenamefont {Das},
  \citenamefont {Sen},\ and\ \citenamefont {Mahapatra}}]{12}%
  \BibitemOpen
  \bibfield  {author} {\bibinfo {author} {\bibfnamefont {B.}~\bibnamefont
  {Das}}, \bibinfo {author} {\bibfnamefont {D.}~\bibnamefont {Sen}},\ and\
  \bibinfo {author} {\bibfnamefont {S.}~\bibnamefont {Mahapatra}},\ }\href
  {https://doi.org/10.1038/s41598-020-63450-5} {\bibfield  {journal} {\bibinfo
  {journal} {Scientific Reports}\ }\textbf {\bibinfo {volume} {10}},\ \bibinfo
  {pages} {6670} (\bibinfo {year} {2020})}\BibitemShut {NoStop}%
\bibitem [{\citenamefont {Fei}\ \emph {et~al.}(2017)\citenamefont {Fei},
  \citenamefont {Palomaki}, \citenamefont {Wu}, \citenamefont {Zhao},
  \citenamefont {Cai}, \citenamefont {Sun}, \citenamefont {Nguyen},
  \citenamefont {Finney}, \citenamefont {Xu},\ and\ \citenamefont
  {Cobden}}]{13}%
  \BibitemOpen
  \bibfield  {author} {\bibinfo {author} {\bibfnamefont {Z.}~\bibnamefont
  {Fei}}, \bibinfo {author} {\bibfnamefont {T.}~\bibnamefont {Palomaki}},
  \bibinfo {author} {\bibfnamefont {S.}~\bibnamefont {Wu}}, \bibinfo {author}
  {\bibfnamefont {W.}~\bibnamefont {Zhao}}, \bibinfo {author} {\bibfnamefont
  {X.}~\bibnamefont {Cai}}, \bibinfo {author} {\bibfnamefont {B.}~\bibnamefont
  {Sun}}, \bibinfo {author} {\bibfnamefont {P.}~\bibnamefont {Nguyen}},
  \bibinfo {author} {\bibfnamefont {J.}~\bibnamefont {Finney}}, \bibinfo
  {author} {\bibfnamefont {X.}~\bibnamefont {Xu}},\ and\ \bibinfo {author}
  {\bibfnamefont {D.~H.}\ \bibnamefont {Cobden}},\ }\href
  {https://doi.org/10.1038/nphys4091} {\bibfield  {journal} {\bibinfo
  {journal} {Nature Physics}\ }\textbf {\bibinfo {volume} {13}},\ \bibinfo
  {pages} {677} (\bibinfo {year} {2017})}\BibitemShut {NoStop}%
\bibitem [{\citenamefont {Zeng}(2024)}]{14}%
  \BibitemOpen
  \bibfield  {author} {\bibinfo {author} {\bibfnamefont {W.}~\bibnamefont
  {Zeng}},\ }\href {https://doi.org/10.1103/PhysRevB.110.205406} {\bibfield
  {journal} {\bibinfo  {journal} {Phys. Rev. B}\ }\textbf {\bibinfo {volume}
  {110}},\ \bibinfo {pages} {205406} (\bibinfo {year} {2024})}\BibitemShut
  {NoStop}%
\bibitem [{\citenamefont {Fatemi}\ \emph {et~al.}(2018)\citenamefont {Fatemi},
  \citenamefont {Wu}, \citenamefont {Cao}, \citenamefont {Bretheau},
  \citenamefont {Gibson}, \citenamefont {Watanabe}, \citenamefont {Taniguchi},
  \citenamefont {Cava},\ and\ \citenamefont {Jarillo-Herrero}}]{15}%
  \BibitemOpen
  \bibfield  {author} {\bibinfo {author} {\bibfnamefont {V.}~\bibnamefont
  {Fatemi}}, \bibinfo {author} {\bibfnamefont {S.}~\bibnamefont {Wu}}, \bibinfo
  {author} {\bibfnamefont {Y.}~\bibnamefont {Cao}}, \bibinfo {author}
  {\bibfnamefont {L.}~\bibnamefont {Bretheau}}, \bibinfo {author}
  {\bibfnamefont {Q.~D.}\ \bibnamefont {Gibson}}, \bibinfo {author}
  {\bibfnamefont {K.}~\bibnamefont {Watanabe}}, \bibinfo {author}
  {\bibfnamefont {T.}~\bibnamefont {Taniguchi}}, \bibinfo {author}
  {\bibfnamefont {R.~J.}\ \bibnamefont {Cava}},\ and\ \bibinfo {author}
  {\bibfnamefont {P.}~\bibnamefont {Jarillo-Herrero}},\ }\href
  {https://doi.org/10.1126/science.aar4642} {\bibfield  {journal} {\bibinfo
  {journal} {Science}\ }\textbf {\bibinfo {volume} {362}},\ \bibinfo {pages}
  {926} (\bibinfo {year} {2018})}\BibitemShut {NoStop}%
\bibitem [{\citenamefont {Lopez-Bezanilla}\ and\ \citenamefont
  {Littlewood}(2016)}]{16}%
  \BibitemOpen
  \bibfield  {author} {\bibinfo {author} {\bibfnamefont {A.}~\bibnamefont
  {Lopez-Bezanilla}}\ and\ \bibinfo {author} {\bibfnamefont {P.~B.}\
  \bibnamefont {Littlewood}},\ }\href
  {https://doi.org/10.1103/PhysRevB.93.241405} {\bibfield  {journal} {\bibinfo
  {journal} {Phys. Rev. B}\ }\textbf {\bibinfo {volume} {93}},\ \bibinfo
  {pages} {241405} (\bibinfo {year} {2016})}\BibitemShut {NoStop}%
\bibitem [{\citenamefont {Zabolotskiy}\ and\ \citenamefont
  {Lozovik}(2016)}]{17}%
  \BibitemOpen
  \bibfield  {author} {\bibinfo {author} {\bibfnamefont {A.~D.}\ \bibnamefont
  {Zabolotskiy}}\ and\ \bibinfo {author} {\bibfnamefont {Y.~E.}\ \bibnamefont
  {Lozovik}},\ }\href {https://doi.org/10.1103/PhysRevB.94.165403} {\bibfield
  {journal} {\bibinfo  {journal} {Phys. Rev. B}\ }\textbf {\bibinfo {volume}
  {94}},\ \bibinfo {pages} {165403} (\bibinfo {year} {2016})}\BibitemShut
  {NoStop}%
\bibitem [{\citenamefont {Li}\ and\ \citenamefont {Chang}(2009)}]{18}%
  \BibitemOpen
  \bibfield  {author} {\bibinfo {author} {\bibfnamefont {J.}~\bibnamefont
  {Li}}\ and\ \bibinfo {author} {\bibfnamefont {K.}~\bibnamefont {Chang}},\
  }\bibfield  {journal} {\bibinfo  {journal} {Applied Physics Letters}\
  }\textbf {\bibinfo {volume} {95}},\ \href {https://doi.org/10.1063/1.3268475}
  {10.1063/1.3268475} (\bibinfo {year} {2009})\BibitemShut {NoStop}%
\bibitem [{\citenamefont {Miao}\ \emph {et~al.}(2012)\citenamefont {Miao},
  \citenamefont {Yan}, \citenamefont {Van~de Walle}, \citenamefont {Lou},
  \citenamefont {Li},\ and\ \citenamefont {Chang}}]{19}%
  \BibitemOpen
  \bibfield  {author} {\bibinfo {author} {\bibfnamefont {M.~S.}\ \bibnamefont
  {Miao}}, \bibinfo {author} {\bibfnamefont {Q.}~\bibnamefont {Yan}}, \bibinfo
  {author} {\bibfnamefont {C.~G.}\ \bibnamefont {Van~de Walle}}, \bibinfo
  {author} {\bibfnamefont {W.~K.}\ \bibnamefont {Lou}}, \bibinfo {author}
  {\bibfnamefont {L.~L.}\ \bibnamefont {Li}},\ and\ \bibinfo {author}
  {\bibfnamefont {K.}~\bibnamefont {Chang}},\ }\href
  {https://doi.org/10.1103/PhysRevLett.109.186803} {\bibfield  {journal}
  {\bibinfo  {journal} {Phys. Rev. Lett.}\ }\textbf {\bibinfo {volume} {109}},\
  \bibinfo {pages} {186803} (\bibinfo {year} {2012})}\BibitemShut {NoStop}%
\bibitem [{\citenamefont {Zhang}\ \emph {et~al.}(2013)\citenamefont {Zhang},
  \citenamefont {Lou}, \citenamefont {Miao}, \citenamefont {Zhang},\ and\
  \citenamefont {Chang}}]{20}%
  \BibitemOpen
  \bibfield  {author} {\bibinfo {author} {\bibfnamefont {D.}~\bibnamefont
  {Zhang}}, \bibinfo {author} {\bibfnamefont {W.}~\bibnamefont {Lou}}, \bibinfo
  {author} {\bibfnamefont {M.}~\bibnamefont {Miao}}, \bibinfo {author}
  {\bibfnamefont {S.-c.}\ \bibnamefont {Zhang}},\ and\ \bibinfo {author}
  {\bibfnamefont {K.}~\bibnamefont {Chang}},\ }\href
  {https://doi.org/10.1103/PhysRevLett.111.156402} {\bibfield  {journal}
  {\bibinfo  {journal} {Phys. Rev. Lett.}\ }\textbf {\bibinfo {volume} {111}},\
  \bibinfo {pages} {156402} (\bibinfo {year} {2013})}\BibitemShut {NoStop}%
\bibitem [{\citenamefont {Ezawa}(2012)}]{21}%
  \BibitemOpen
  \bibfield  {author} {\bibinfo {author} {\bibfnamefont {M.}~\bibnamefont
  {Ezawa}},\ }\href {https://doi.org/10.1103/PhysRevLett.109.055502} {\bibfield
   {journal} {\bibinfo  {journal} {Phys. Rev. Lett.}\ }\textbf {\bibinfo
  {volume} {109}},\ \bibinfo {pages} {055502} (\bibinfo {year}
  {2012})}\BibitemShut {NoStop}%
\bibitem [{\citenamefont {Tan}\ \emph {et~al.}(2021)\citenamefont {Tan},
  \citenamefont {Yan}, \citenamefont {Zhao}, \citenamefont {Guo},\ and\
  \citenamefont {Chang}}]{22}%
  \BibitemOpen
  \bibfield  {author} {\bibinfo {author} {\bibfnamefont {C.-Y.}\ \bibnamefont
  {Tan}}, \bibinfo {author} {\bibfnamefont {C.-X.}\ \bibnamefont {Yan}},
  \bibinfo {author} {\bibfnamefont {Y.-H.}\ \bibnamefont {Zhao}}, \bibinfo
  {author} {\bibfnamefont {H.}~\bibnamefont {Guo}},\ and\ \bibinfo {author}
  {\bibfnamefont {H.-R.}\ \bibnamefont {Chang}},\ }\href
  {https://doi.org/10.1103/PhysRevB.103.125425} {\bibfield  {journal} {\bibinfo
   {journal} {Phys. Rev. B}\ }\textbf {\bibinfo {volume} {103}},\ \bibinfo
  {pages} {125425} (\bibinfo {year} {2021})}\BibitemShut {NoStop}%
\bibitem [{\citenamefont {Balassis}\ \emph {et~al.}(2022)\citenamefont
  {Balassis}, \citenamefont {Gumbs},\ and\ \citenamefont {Roslyak}}]{23}%
  \BibitemOpen
  \bibfield  {author} {\bibinfo {author} {\bibfnamefont {A.}~\bibnamefont
  {Balassis}}, \bibinfo {author} {\bibfnamefont {G.}~\bibnamefont {Gumbs}},\
  and\ \bibinfo {author} {\bibfnamefont {O.}~\bibnamefont {Roslyak}},\ }\href
  {https://doi.org/https://doi.org/10.1016/j.physleta.2022.128353} {\bibfield
  {journal} {\bibinfo  {journal} {Physics Letters A}\ }\textbf {\bibinfo
  {volume} {449}},\ \bibinfo {pages} {128353} (\bibinfo {year}
  {2022})}\BibitemShut {NoStop}%
\bibitem [{\citenamefont {Wang}\ \emph {et~al.}(2023)\citenamefont {Wang},
  \citenamefont {Hu}, \citenamefont {Lv},\ and\ \citenamefont {Jin}}]{24}%
  \BibitemOpen
  \bibfield  {author} {\bibinfo {author} {\bibfnamefont {D.}~\bibnamefont
  {Wang}}, \bibinfo {author} {\bibfnamefont {A.}~\bibnamefont {Hu}}, \bibinfo
  {author} {\bibfnamefont {J.-P.}\ \bibnamefont {Lv}},\ and\ \bibinfo {author}
  {\bibfnamefont {G.}~\bibnamefont {Jin}},\ }\href
  {https://doi.org/10.1103/PhysRevB.107.035301} {\bibfield  {journal} {\bibinfo
   {journal} {Phys. Rev. B}\ }\textbf {\bibinfo {volume} {107}},\ \bibinfo
  {pages} {035301} (\bibinfo {year} {2023})}\BibitemShut {NoStop}%
\bibitem [{\citenamefont {Weiss}\ \emph {et~al.}(1989)\citenamefont {Weiss},
  \citenamefont {Klitzing}, \citenamefont {Ploog},\ and\ \citenamefont
  {Weimann}}]{25}%
  \BibitemOpen
  \bibfield  {author} {\bibinfo {author} {\bibfnamefont {D.}~\bibnamefont
  {Weiss}}, \bibinfo {author} {\bibfnamefont {K.}~\bibnamefont {Klitzing}},
  \bibinfo {author} {\bibfnamefont {K.}~\bibnamefont {Ploog}},\ and\ \bibinfo
  {author} {\bibfnamefont {G.}~\bibnamefont {Weimann}},\ }\href
  {https://doi.org/10.1209/0295-5075/8/2/012} {\bibfield  {journal} {\bibinfo
  {journal} {Europhysics Letters}\ }\textbf {\bibinfo {volume} {8}},\ \bibinfo
  {pages} {179} (\bibinfo {year} {1989})}\BibitemShut {NoStop}%
\bibitem [{\citenamefont {Gerhardts}\ \emph
  {et~al.}(1989{\natexlab{a}})\citenamefont {Gerhardts}, \citenamefont
  {Weiss},\ and\ \citenamefont {Klitzing}}]{26}%
  \BibitemOpen
  \bibfield  {author} {\bibinfo {author} {\bibfnamefont {R.~R.}\ \bibnamefont
  {Gerhardts}}, \bibinfo {author} {\bibfnamefont {D.}~\bibnamefont {Weiss}},\
  and\ \bibinfo {author} {\bibfnamefont {K.~v.}\ \bibnamefont {Klitzing}},\
  }\href {https://doi.org/10.1103/PhysRevLett.62.1173} {\bibfield  {journal}
  {\bibinfo  {journal} {Phys. Rev. Lett.}\ }\textbf {\bibinfo {volume} {62}},\
  \bibinfo {pages} {1173} (\bibinfo {year} {1989}{\natexlab{a}})}\BibitemShut
  {NoStop}%
\bibitem [{\citenamefont {Winkler}\ \emph {et~al.}(1989)\citenamefont
  {Winkler}, \citenamefont {Kotthaus},\ and\ \citenamefont {Ploog}}]{27}%
  \BibitemOpen
  \bibfield  {author} {\bibinfo {author} {\bibfnamefont {R.~W.}\ \bibnamefont
  {Winkler}}, \bibinfo {author} {\bibfnamefont {J.~P.}\ \bibnamefont
  {Kotthaus}},\ and\ \bibinfo {author} {\bibfnamefont {K.}~\bibnamefont
  {Ploog}},\ }\href {https://doi.org/10.1103/PhysRevLett.62.1177} {\bibfield
  {journal} {\bibinfo  {journal} {Phys. Rev. Lett.}\ }\textbf {\bibinfo
  {volume} {62}},\ \bibinfo {pages} {1177} (\bibinfo {year}
  {1989})}\BibitemShut {NoStop}%
\bibitem [{\citenamefont {Izawa}\ \emph {et~al.}(1995)\citenamefont {Izawa},
  \citenamefont {Katsumoto}, \citenamefont {Endo},\ and\ \citenamefont
  {Iye}}]{28}%
  \BibitemOpen
  \bibfield  {author} {\bibinfo {author} {\bibfnamefont {S.-i.}\ \bibnamefont
  {Izawa}}, \bibinfo {author} {\bibfnamefont {S.}~\bibnamefont {Katsumoto}},
  \bibinfo {author} {\bibfnamefont {A.}~\bibnamefont {Endo}},\ and\ \bibinfo
  {author} {\bibfnamefont {Y.}~\bibnamefont {Iye}},\ }\href
  {https://doi.org/10.1143/JPSJ.64.706} {\bibfield  {journal} {\bibinfo
  {journal} {Journal of the Physical Society of Japan}\ }\textbf {\bibinfo
  {volume} {64}},\ \bibinfo {pages} {706} (\bibinfo {year} {1995})}\BibitemShut
  {NoStop}%
\bibitem [{\citenamefont {Carmona}\ \emph {et~al.}(1995)\citenamefont
  {Carmona}, \citenamefont {Geim}, \citenamefont {Nogaret}, \citenamefont
  {Main}, \citenamefont {Foster}, \citenamefont {Henini}, \citenamefont
  {Beaumont},\ and\ \citenamefont {Blamire}}]{29}%
  \BibitemOpen
  \bibfield  {author} {\bibinfo {author} {\bibfnamefont {H.~A.}\ \bibnamefont
  {Carmona}}, \bibinfo {author} {\bibfnamefont {A.~K.}\ \bibnamefont {Geim}},
  \bibinfo {author} {\bibfnamefont {A.}~\bibnamefont {Nogaret}}, \bibinfo
  {author} {\bibfnamefont {P.~C.}\ \bibnamefont {Main}}, \bibinfo {author}
  {\bibfnamefont {T.~J.}\ \bibnamefont {Foster}}, \bibinfo {author}
  {\bibfnamefont {M.}~\bibnamefont {Henini}}, \bibinfo {author} {\bibfnamefont
  {S.~P.}\ \bibnamefont {Beaumont}},\ and\ \bibinfo {author} {\bibfnamefont
  {M.~G.}\ \bibnamefont {Blamire}},\ }\href
  {https://doi.org/10.1103/PhysRevLett.74.3009} {\bibfield  {journal} {\bibinfo
   {journal} {Phys. Rev. Lett.}\ }\textbf {\bibinfo {volume} {74}},\ \bibinfo
  {pages} {3009} (\bibinfo {year} {1995})}\BibitemShut {NoStop}%
\bibitem [{\citenamefont {Ye}\ \emph {et~al.}(1995)\citenamefont {Ye},
  \citenamefont {Weiss}, \citenamefont {Gerhardts}, \citenamefont {Seeger},
  \citenamefont {von Klitzing}, \citenamefont {Eberl},\ and\ \citenamefont
  {Nickel}}]{30}%
  \BibitemOpen
  \bibfield  {author} {\bibinfo {author} {\bibfnamefont {P.~D.}\ \bibnamefont
  {Ye}}, \bibinfo {author} {\bibfnamefont {D.}~\bibnamefont {Weiss}}, \bibinfo
  {author} {\bibfnamefont {R.~R.}\ \bibnamefont {Gerhardts}}, \bibinfo {author}
  {\bibfnamefont {M.}~\bibnamefont {Seeger}}, \bibinfo {author} {\bibfnamefont
  {K.}~\bibnamefont {von Klitzing}}, \bibinfo {author} {\bibfnamefont
  {K.}~\bibnamefont {Eberl}},\ and\ \bibinfo {author} {\bibfnamefont
  {H.}~\bibnamefont {Nickel}},\ }\href
  {https://doi.org/10.1103/PhysRevLett.74.3013} {\bibfield  {journal} {\bibinfo
   {journal} {Phys. Rev. Lett.}\ }\textbf {\bibinfo {volume} {74}},\ \bibinfo
  {pages} {3013} (\bibinfo {year} {1995})}\BibitemShut {NoStop}%
\bibitem [{\citenamefont {Gerhardts}\ \emph
  {et~al.}(1989{\natexlab{b}})\citenamefont {Gerhardts}, \citenamefont
  {Weiss},\ and\ \citenamefont {Klitzing}}]{31}%
  \BibitemOpen
  \bibfield  {author} {\bibinfo {author} {\bibfnamefont {R.~R.}\ \bibnamefont
  {Gerhardts}}, \bibinfo {author} {\bibfnamefont {D.}~\bibnamefont {Weiss}},\
  and\ \bibinfo {author} {\bibfnamefont {K.~v.}\ \bibnamefont {Klitzing}},\
  }\href {https://doi.org/10.1103/PhysRevLett.62.1173} {\bibfield  {journal}
  {\bibinfo  {journal} {Phys. Rev. Lett.}\ }\textbf {\bibinfo {volume} {62}},\
  \bibinfo {pages} {1173} (\bibinfo {year} {1989}{\natexlab{b}})}\BibitemShut
  {NoStop}%
\bibitem [{\citenamefont {Matulis}\ and\ \citenamefont {Peeters}(2007)}]{32}%
  \BibitemOpen
  \bibfield  {author} {\bibinfo {author} {\bibfnamefont {A.}~\bibnamefont
  {Matulis}}\ and\ \bibinfo {author} {\bibfnamefont {F.~M.}\ \bibnamefont
  {Peeters}},\ }\href {https://doi.org/10.1103/PhysRevB.75.125429} {\bibfield
  {journal} {\bibinfo  {journal} {Phys. Rev. B}\ }\textbf {\bibinfo {volume}
  {75}},\ \bibinfo {pages} {125429} (\bibinfo {year} {2007})}\BibitemShut
  {NoStop}%
\bibitem [{\citenamefont {Tahir}\ and\ \citenamefont {Sabeeh}(2008)}]{33}%
  \BibitemOpen
  \bibfield  {author} {\bibinfo {author} {\bibfnamefont {M.}~\bibnamefont
  {Tahir}}\ and\ \bibinfo {author} {\bibfnamefont {K.}~\bibnamefont {Sabeeh}},\
  }\href {https://doi.org/10.1103/PhysRevB.77.195421} {\bibfield  {journal}
  {\bibinfo  {journal} {Phys. Rev. B}\ }\textbf {\bibinfo {volume} {77}},\
  \bibinfo {pages} {195421} (\bibinfo {year} {2008})}\BibitemShut {NoStop}%
\bibitem [{\citenamefont {Zarenia}\ \emph {et~al.}(2012)\citenamefont
  {Zarenia}, \citenamefont {Vasilopoulos},\ and\ \citenamefont {Peeters}}]{34}%
  \BibitemOpen
  \bibfield  {author} {\bibinfo {author} {\bibfnamefont {M.}~\bibnamefont
  {Zarenia}}, \bibinfo {author} {\bibfnamefont {P.}~\bibnamefont
  {Vasilopoulos}},\ and\ \bibinfo {author} {\bibfnamefont {F.~M.}\ \bibnamefont
  {Peeters}},\ }\href {https://doi.org/10.1103/PhysRevB.85.245426} {\bibfield
  {journal} {\bibinfo  {journal} {Phys. Rev. B}\ }\textbf {\bibinfo {volume}
  {85}},\ \bibinfo {pages} {245426} (\bibinfo {year} {2012})}\BibitemShut
  {NoStop}%
\bibitem [{\citenamefont {Islam}\ and\ \citenamefont {Dutta}(2017)}]{35}%
  \BibitemOpen
  \bibfield  {author} {\bibinfo {author} {\bibfnamefont {S.~F.}\ \bibnamefont
  {Islam}}\ and\ \bibinfo {author} {\bibfnamefont {P.}~\bibnamefont {Dutta}},\
  }\href {https://doi.org/10.1103/PhysRevB.96.045418} {\bibfield  {journal}
  {\bibinfo  {journal} {Phys. Rev. B}\ }\textbf {\bibinfo {volume} {96}},\
  \bibinfo {pages} {045418} (\bibinfo {year} {2017})}\BibitemShut {NoStop}%
\bibitem [{\citenamefont {Tahir}\ and\ \citenamefont
  {Vasilopoulos}(2017)}]{36}%
  \BibitemOpen
  \bibfield  {author} {\bibinfo {author} {\bibfnamefont {M.}~\bibnamefont
  {Tahir}}\ and\ \bibinfo {author} {\bibfnamefont {P.}~\bibnamefont
  {Vasilopoulos}},\ }\href {https://doi.org/10.1088/1361-648X/aa8428}
  {\bibfield  {journal} {\bibinfo  {journal} {Journal of Physics: Condensed
  Matter}\ }\textbf {\bibinfo {volume} {29}},\ \bibinfo {pages} {425302}
  (\bibinfo {year} {2017})}\BibitemShut {NoStop}%
\bibitem [{\citenamefont {Wang}\ \emph {et~al.}(2005)\citenamefont {Wang},
  \citenamefont {Vasilopoulos},\ and\ \citenamefont {Peeters}}]{37}%
  \BibitemOpen
  \bibfield  {author} {\bibinfo {author} {\bibfnamefont {X.~F.}\ \bibnamefont
  {Wang}}, \bibinfo {author} {\bibfnamefont {P.}~\bibnamefont {Vasilopoulos}},\
  and\ \bibinfo {author} {\bibfnamefont {F.~M.}\ \bibnamefont {Peeters}},\
  }\href {https://doi.org/10.1103/PhysRevB.71.125301} {\bibfield  {journal}
  {\bibinfo  {journal} {Phys. Rev. B}\ }\textbf {\bibinfo {volume} {71}},\
  \bibinfo {pages} {125301} (\bibinfo {year} {2005})}\BibitemShut {NoStop}%
\bibitem [{\citenamefont {Islam}\ and\ \citenamefont {Ghosh}(2012)}]{38}%
  \BibitemOpen
  \bibfield  {author} {\bibinfo {author} {\bibfnamefont {S.~F.}\ \bibnamefont
  {Islam}}\ and\ \bibinfo {author} {\bibfnamefont {T.~K.}\ \bibnamefont
  {Ghosh}},\ }\href {https://doi.org/10.1088/0953-8984/24/18/185303} {\bibfield
   {journal} {\bibinfo  {journal} {Journal of Physics: Condensed Matter}\
  }\textbf {\bibinfo {volume} {24}},\ \bibinfo {pages} {185303} (\bibinfo
  {year} {2012})}\BibitemShut {NoStop}%
\bibitem [{\citenamefont {Islam}\ and\ \citenamefont {Ghosh}(2014)}]{39}%
  \BibitemOpen
  \bibfield  {author} {\bibinfo {author} {\bibfnamefont {S.~F.}\ \bibnamefont
  {Islam}}\ and\ \bibinfo {author} {\bibfnamefont {T.~K.}\ \bibnamefont
  {Ghosh}},\ }\href {https://doi.org/10.1088/0953-8984/26/33/335303} {\bibfield
   {journal} {\bibinfo  {journal} {Journal of Physics: Condensed Matter}\
  }\textbf {\bibinfo {volume} {26}},\ \bibinfo {pages} {335303} (\bibinfo
  {year} {2014})}\BibitemShut {NoStop}%
\bibitem [{\citenamefont {Shakouri}\ \emph {et~al.}(2014)\citenamefont
  {Shakouri}, \citenamefont {Vasilopoulos}, \citenamefont {Vargiamidis},\ and\
  \citenamefont {Peeters}}]{40}%
  \BibitemOpen
  \bibfield  {author} {\bibinfo {author} {\bibfnamefont {K.}~\bibnamefont
  {Shakouri}}, \bibinfo {author} {\bibfnamefont {P.}~\bibnamefont
  {Vasilopoulos}}, \bibinfo {author} {\bibfnamefont {V.}~\bibnamefont
  {Vargiamidis}},\ and\ \bibinfo {author} {\bibfnamefont {F.~M.}\ \bibnamefont
  {Peeters}},\ }\href {https://doi.org/10.1103/PhysRevB.90.125444} {\bibfield
  {journal} {\bibinfo  {journal} {Phys. Rev. B}\ }\textbf {\bibinfo {volume}
  {90}},\ \bibinfo {pages} {125444} (\bibinfo {year} {2014})}\BibitemShut
  {NoStop}%
\bibitem [{\citenamefont {Islam}\ and\ \citenamefont {Jayannavar}(2017)}]{41}%
  \BibitemOpen
  \bibfield  {author} {\bibinfo {author} {\bibfnamefont {S.~F.}\ \bibnamefont
  {Islam}}\ and\ \bibinfo {author} {\bibfnamefont {A.~M.}\ \bibnamefont
  {Jayannavar}},\ }\href {https://doi.org/10.1103/PhysRevB.96.235405}
  {\bibfield  {journal} {\bibinfo  {journal} {Phys. Rev. B}\ }\textbf {\bibinfo
  {volume} {96}},\ \bibinfo {pages} {235405} (\bibinfo {year}
  {2017})}\BibitemShut {NoStop}%
\bibitem [{\citenamefont {Peres}\ and\ \citenamefont {Castro}(2007)}]{42}%
  \BibitemOpen
  \bibfield  {author} {\bibinfo {author} {\bibfnamefont {N.~M.~R.}\
  \bibnamefont {Peres}}\ and\ \bibinfo {author} {\bibfnamefont {E.~V.}\
  \bibnamefont {Castro}},\ }\href
  {https://doi.org/10.1088/0953-8984/19/40/406231} {\bibfield  {journal}
  {\bibinfo  {journal} {Journal of Physics: Condensed Matter}\ }\textbf
  {\bibinfo {volume} {19}},\ \bibinfo {pages} {406231} (\bibinfo {year}
  {2007})}\BibitemShut {NoStop}%
\bibitem [{\citenamefont {Kubo}\ \emph {et~al.}(1959)\citenamefont {Kubo},
  \citenamefont {Hasegawa},\ and\ \citenamefont {Hashitsume}}]{43}%
  \BibitemOpen
  \bibfield  {author} {\bibinfo {author} {\bibfnamefont {R.}~\bibnamefont
  {Kubo}}, \bibinfo {author} {\bibfnamefont {H.}~\bibnamefont {Hasegawa}},\
  and\ \bibinfo {author} {\bibfnamefont {N.}~\bibnamefont {Hashitsume}},\
  }\href {https://doi.org/10.1143/JPSJ.14.56} {\bibfield  {journal} {\bibinfo
  {journal} {Journal of the Physical Society of Japan}\ }\textbf {\bibinfo
  {volume} {14}},\ \bibinfo {pages} {56} (\bibinfo {year} {1959})}\BibitemShut
  {NoStop}%
\bibitem [{\citenamefont {Peeters}\ and\ \citenamefont
  {Vasilopoulos}(1992)}]{44}%
  \BibitemOpen
  \bibfield  {author} {\bibinfo {author} {\bibfnamefont {F.~M.}\ \bibnamefont
  {Peeters}}\ and\ \bibinfo {author} {\bibfnamefont {P.}~\bibnamefont
  {Vasilopoulos}},\ }\href {https://doi.org/10.1103/PhysRevB.46.4667}
  {\bibfield  {journal} {\bibinfo  {journal} {Phys. Rev. B}\ }\textbf {\bibinfo
  {volume} {46}},\ \bibinfo {pages} {4667} (\bibinfo {year}
  {1992})}\BibitemShut {NoStop}%
\bibitem [{\citenamefont {Charbonneau}\ \emph {et~al.}(1982)\citenamefont
  {Charbonneau}, \citenamefont {Van~Vliet},\ and\ \citenamefont
  {Vasilopoulos}}]{45}%
  \BibitemOpen
  \bibfield  {author} {\bibinfo {author} {\bibfnamefont {M.}~\bibnamefont
  {Charbonneau}}, \bibinfo {author} {\bibfnamefont {K.}~\bibnamefont
  {Van~Vliet}},\ and\ \bibinfo {author} {\bibfnamefont {P.}~\bibnamefont
  {Vasilopoulos}},\ }\href {https://doi.org/10.1063/1.525355} {\bibfield
  {journal} {\bibinfo  {journal} {Journal of Mathematical Physics}\ }\textbf
  {\bibinfo {volume} {23}},\ \bibinfo {pages} {318} (\bibinfo {year}
  {1982})}\BibitemShut {NoStop}%
\end{thebibliography}%

\end{document}